%% file: Broadcast_Games.tex
\documentclass[10pt,twocolumn]{IEEEtran}
\usepackage[T1]{fontenc}
\usepackage{amsmath,amssymb}
\usepackage{graphicx}
\usepackage{epsfig}
\usepackage{cite}
\usepackage{setspace}
\usepackage{color}
\usepackage{enumerate}
\usepackage{balance}

\newcommand{\mA}{\mathcal A}

\newcommand{\tQ}{\tilde Q}
\newcommand{\hQ}{\hat Q}
\newcommand{\tL}{\tilde L}
\newcommand{\hL}{\hat L}

\newcommand{\mT}{\mathcal T}
\newcommand{\mN}{\mathcal N}
\newcommand{\mQ}{\mathcal Q}
\newcommand{\tl}{{\tilde \lambda}}
\newcommand{\hl}{{\hat \lambda}}
\newcommand{\ur}{\underline{r}}

\newtheorem{thm}{Theorem}
\newtheorem{mydef}{Definition}
\newtheorem{lemma}{Lemma}
\newtheorem{example}{Example}
\newtheorem{prop}{Proposition}

\title{Broadcast Channel Games: Equilibrium Characterization and a MIMO MAC-BC Game Duality}
\author{Srinivas Yerramalli, Rahul Jain and Urbashi Mitra \\
       University of Southern California \\
      \{srinivas.yerramalli,rahul.jain,ubli\}@usc.edu
      \thanks{This research has been funded in part by the following grants and organizations: NSF CCF-0917343, NSF IIS 0917410, CAREER CNS-0954116, ONR N00014-09-1-0700, NSF CNS 0832186, NSF CCF-1117896 and NSF CNS-082175.}}

\begin{document}
\maketitle

\vspace{-0.5in}
\begin{abstract}
\input{abstract.tex}

\end{abstract}

\section{Introduction}
\label{sec:intro}
\input{intro.tex}

\section{Preliminaries}
\label{sec:preliminaries}
\input{preliminaries.tex}

\section{The Broadcast Channel as a Generalized Nash Equilibrium Problem}
\label{sec:game_model}
\input{game_model.tex}

\section{Relationship between NoEs and Pareto Optimal Solutions}
\label{sec:relationship}
\input{relationship.tex}

\section{Game Theoretic Duality between MAC and BC}
\label{sec:duality}
\input{duality.tex}

\section{Conclusions}
\label{sec:conclusions}
\input{conclusions.tex}


\bibliographystyle{IEEEtran}
\bibliography{references}

\balance

\end{document}

%% file: abstract.tex
The emergence of heterogeneous decentralized networks without a central controller, such as device-to-device communication systems, has created the need for new problem frameworks to design and analyze the performance of such networks. As a key step towards such an analysis for general networks, this paper examines the strategic behavior of \emph{receivers} in a Gaussian broadcast channel (BC) and \emph{transmitters} in a multiple access channel (MAC) with sum power constraints (sum power MAC) using the framework of non-cooperative game theory. These signaling scenarios are modeled as generalized Nash equilibrium problems (GNEPs) with jointly convex and coupled constraints and the existence and uniqueness of equilibrium achieving strategies and equilibrium utilities are characterized for both the Gaussian BC and the sum power MAC. The relationship between Pareto-optimal boundary points of the capacity region and the generalized Nash equilibria (GNEs) are derived for the several special cases and in all these cases it is shown that all the GNEs are Pareto-optimal, demonstrating that there is no loss in efficiency when players adopt strategic behavior in these scenarios. Several key equivalence relations are derived and used to demonstrate a game-theoretic duality between the Gaussian MAC and the Gaussian BC. This duality allows a parametrized computation of the equilibria of the BC in terms of the equilibria of the MAC and paves the way to translate several MAC results to the dual BC scenario.

%% file: intro.tex
Several wireless and cellular networks currently in operation are centrally controlled by an operator who is assumed to know the state of the network and operate it in an optimal fashion to maximize throughput and minimize interference in a fair and efficient manner. However, due to the recent emergence of heterogeneous and decentralized wireless networks, such as device-to-device communication systems, in which several nodes are owned by different operators, it can no longer be assumed that there exists a central controller which can optimize the performance of a network as a whole. Thus new problem frameworks are needed to both design and analyze the performance of such networks. In recent literature, game theory has been used extensively to model, understand and drive the interactions between nodes in heterogeneous networks which act to optimize their own individual objectives. For example, several aspects of the multiple access channel (MAC) and the interference channel (IC) have been extensively studied using non-cooperative game theory (see \cite{Scutari10,survey11} for a detailed survey).

In a typical non-cooperative game, the choice of actions of a player affects the \emph{utility} obtained by \textit{every} player, but does not change the set of available \emph{actions} for other players. For example, the Gaussian MAC, the Gaussian IC and its variants belong to this category of games \cite{La04,Lai08,Belmega10,Srinivas11ISIT}. However, in some scenarios, such as the Gaussian broadcast channel (BC), in which the \emph{receivers} can be considered the players in a game, the choice of actions of all the players is jointly constrained by a transmit power or covariance constraint as the choice of a transmit covariance matrix for one receiver will constrain the choice of transmit covariance matrices for the other receivers. In \cite{Pang08intfchannel}, the problem of maintaining a minimum rate over parallel Gaussian ICs, subject to a sum power constraint for each user, is considered. The choice of power allocation of a player for a given channel is then influenced by the choice of power allocation of other players on other interference channels to maintain an overall desired rate and hence is also a game with coupled constraints on the strategies of adopted by the players. Such joint constraints on the strategies of the players result in the feasible set of each player being a function of the choice of strategies of the other players. Interaction of players at the level of feasible sets makes the analysis of such games more challenging than standard non-cooperative games. The problem of determining the equilibria of games with coupled constraints is called a \emph{generalized Nash equilibrium problem (GNEP)} \cite{Scutari10,Pang08intfchannel,Rosen65} and the points themselves are called generalized Nash equilibria (GNE).

From a game theoretic perspective, the BC and its related problems have received little attention relative to other channels such as the MAC and the IC. A discrete memoryless BC with $2$ users and a resource manager was considered in \cite{Su08} and the impact of the information available to the resource manager in modifying the utility of each user is studied. Our previous work \cite{Srinivas13MAC} considered the device-to-device cooperation problem for the uplink with perfect cooperation and examined the stability of the grand coalition of transmitters in a MAC. The work in \cite{Pantisano2012spectrum} proposes a framework for macrocell-femtocell cooperation to delegate transmission in which the femtocell acts as a relay for the macro cell and considers coalition formation for this scenario. In \cite{Xu12}, interference aware resource allocation for device-to-device communication is considered and a sequential second price auction is proposed for sharing the spectrum and a mode selection algorithm for communication based on coalitional games is proposed in \cite{Akkarajitsakul2012mode}. Several works in the literature has used game theory as an effective tool  to study various aspects of device-to-device cooperation. However, we observe there is a significant need to introduce new game theoretic models to analyze complex networks. To this end, we introduce a game theoretic model for the BC which enables the study of downlink cooperation with rational and selfish players. Such a model could also pave the way for analysis of cognitive radio scenarios in which the primary and multiple secondary users can operate simultaneously on the same spectrum \cite{Zhang2009weighted}.

In this paper, we propose to use the framework of games with coupled constraints to characterize the equilibria of the MIMO Gaussian BC and the MIMO Gaussian sum power MAC. Our goal in this paper is to analyze the Gaussian BC and sum power MAC from a game-theoretic perspective and interpret several achievable rates/capacity results \cite{Weingarten06} for the BC in this framework. Our work shows the existence of GNEs for the general Gaussian BC and characterizes the uniqueness of equilibrium utilities for the general case and equilibrium achieving strategies for a special class of BCs termed aligned and degraded broadcast channels (ADBCs) \cite{Weingarten06}. We then consider the MIMO MAC with sum power constraints (henceforth called the \emph{sum power MAC}). The sum power MAC is a problem that is closely related to the BC by an information-theoretic duality and \cite{Vishwanath03} shows that the capacity region for the sum power MAC is the same as the capacity region for an equivalently defined BC. We characterize the existence and uniqueness of equilibria for the sum power MAC and show that while the utility at equilibrium, given a decoding order, is the same for all the players, the equilibrium achieving strategies may not necessarily be unique. By deriving a relation between Pareto-optimal points and equilibrium rates we show that every rate point on the Pareto-optimal boundary of the BC and the sum power MAC is an equilibrium rate point. 

In information theory, owing to the structure of the MIMO BC, associated optimization problems such as capacity region computation and beamforming optimization are typically highly non-convex problems and cannot be solved directly. One feasible approach that has proved effective in the literature is to transform the non-convex BC problems into a convex dual MAC problem, which is easier to deal with and then transform the solution back into the domain of the BC \cite{Vishwanath03} (see \cite{Zhang2012gaussian} for a comprehensive survey on the general duality between MAC and BC with possibly multiple constraints). The conventional information theoretic duality between the BC and the MAC has been established for several scenarios and has been used to solve several problems of interest to the wireless community. In this paper, we derive a game-theoretic duality between the MAC and the BC. We first develop several equivalence relations between Nash equilibria (NE) of a class of Nash equilibrium problems (NEP) related to the MAC and GNEPs related to the sum power MAC and the BC and then demonstrate a game-theoretic duality between the MAC and the BC. On the lines of the information theoretic duality for the sum power MAC and BC, the game theoretic duality states that every equilibrium point of the BC can be transformed into equilibrium points of the sum power MAC and vice versa. The duality provides us a method to parametrically obtain the GNEs of the BC by solving the NE of the dual MAC and then transform the solution to the BC using \cite{Vishwanath03}. In general, the duality will enable the analysis of BC games by considering equivalent MAC games and allow us to translate key results from the MAC to the BC. 

The rest of the paper is organized as follows. Section \ref{sec:preliminaries} introduces several concepts related to GNEPs and describes various ways of analyzing GNEPs. In addition, Section \ref{sec:preliminaries} derives the condition for the uniqueness of equilibrium achieving strategies for a parameterized GNEP characterization. Using the results in Section \ref{sec:preliminaries}, the existence and uniqueness of equilibria for the BC and the sum power MAC are analyzed in Section \ref{sec:game_model} and Section \ref{sec:sumpowerMAC} respectively. The relationship between Pareto-optimal solutions and equilibria is derived in Section \ref{sec:relationship} and the correspondence between the two solutions is discussed with applicability to both the sum power MAC and the BC. Section \ref{sec:duality} derives several equivalence relations between the MAC and the BC and then shows the existence of a game theoretic duality between the two channels and Section \ref{sec:conclusions} concludes this paper.

%% file: preliminaries.tex
We begin by reviewing several game theoretic concepts for GNEPs. We first introduce the GNEP formulation, the concepts of generalized and normalized equilibria and discuss the relationship between them. Using the formulation in \cite{Rosen65}, we then derive a sufficient condition for the existence and uniqueness of normalized equilibria for GNEPs.

\subsection{Generalized Nash Equilibrium Problems}
Formally, a GNEP consists of $K$ players with each player controlling the variable $Q_k$. We denote by $\mQ$, the vector formed by all these decision variables $ \mQ = (Q_1,Q_2,...,Q_K) $, and by $Q_{-k}$ the vector formed by the decision variables of all other players except the $kth$ player. Each player has an objective function $v_k$ that depends on both his own variables $Q_k$ and the controlling variables of all other players $Q_{-k}$. This function
is called the \emph{utility function} of the $kth$ player and is formally denoted as $v_k(Q_k,Q_{-k})$ or $v_k(\mQ), ~ \mQ = (Q_1,Q_2,...,Q_K)$ to emphasize the dependence on the controlling variables. Furthermore, each player's action must belong to a set $ \mA_k(Q_{-k}) $ that depends on the rival players' actions and that we call the \emph{feasible set} or \emph{action space} of player $k$. We emphasize that the set $\mA_k(Q_{-k})$ is a function of the strategies of the other players. The aim of the $kth$ player, given the actions of all the other players $Q_{-k}$, is to pick a strategy that solves the maximization problem
\begin{equation}
\max_{Q_k} v_k(Q_k,Q_{-k}) ~ \textrm{subject to} ~ Q_k \in \mA_k(Q_{-k}).
\label{e:GNEP}
\end{equation}
Let $ \Psi_k (Q_{-k}) $ denote the set of all the solutions to the $kth$ player's maximization problem given $Q_{-k}$. The GNEP is the problem of finding $Q^{*}_{k}$ such that $$ Q^{*}_k \in \Psi(Q^{*}_{-k}) ~ \textrm{for all} ~ k = 1,2,...,K.$$ Such as point is called a \emph{generalized Nash equilibrium} (GNE) or more generally a solution to the GNEP. A point $Q^{*}$ is therefore an equilibrium, if no player can improve this objective function by changing \emph{unilaterally} to any other point in his feasible set. If we denote by $ \Psi(Q)$ the set $ \Psi(Q) := \times_{i=1}^{K} \Psi_i(Q_{-i}) $ we see that $ Q^{*} $ is a GNE if and only if $ Q^{*} \in \Psi(Q^{*}) $, \textit{i.e.,} if and only if $Q^{*}$ is a \emph{fixed point} of the mapping $ \Psi $. If the feasible set of each player is independent of the actions adopted by all the other players, then the GNEP reduces to the well known Nash equilibrium problem.

In a GNEP, the $kth$ player must know the strategy of the other players to determine his own feasible set, however the other players need to know the strategy of the $kth$ player to determine their own strategy. Thus there is a practical challenge of a game where the players make their choices \emph{simultaneously} and it happens that the consraints are satisfied. However, this view of GNEPs is limited as observed in \cite{Facchinei07} and undervalues the (1) descriptive power of the GNEP model and (2) possibility of using GNEPs to develop rules and protocols, set taxes \textit{etc}. in order to achieve performance goals (see \cite{Facchinei07} for a detailed discussion).

\emph{Remark:} We note that GNEPs have sometimes been used in a normative way in the literature. No one is really playing a game; rather a single decision maker establishes that the outcome of a GNEP is desirable and therefore implements this solution. However, adopting a pricing and penalty based approach (discussed later in this section) to solving GNEPs has been successfully adopted in several works: the power allocation a problem in a Gaussian interference channel \cite{Pang08intfchannel}, design for cognitive radio systems with interference constraints \cite{Pang10CR} \textit{etc}.  

A GNEP usually has multiple or, in many cases, \emph{uncountably} many equilibria \cite{Rosen65,Facchinei07}. While characterizing the GNEs for a general GNEP has proven to be a challenging problem in general, special classes of GNEPs, such as jointly convex scenarios, have been examined in literature and are briefly illustrated below \cite{Rosen65}.

\begin{mydef}
A GNEP is said to be \emph{concave} (convex) if for every player $k$ and every $Q_{-k}$, the utility function $v_k(Q_k,Q_{-k})$ is concave (convex) and the set $ \Psi_k(Q_{-k}) $ is closed and convex.
\end{mydef}

\begin{mydef}
A GNEP is said to be \emph{jointly convex} if the GNEP is concave (convex) and for some closed set $\mQ$ and all $k=1,2,...,K$, we have
\begin{equation}
\Psi_k(Q_{-k}) = \left \{ Q_k | (Q_k,Q_{-k}) \in \mQ \right \}.
\end{equation}
\label{def:jointlyconvex}
\end{mydef}
This class of problems was first studied in detail in a seminal paper by Rosen \cite{Rosen65} and has been referred to as \emph{GNEPs with coupled constraints} or in general \emph{jointly convex GNEPs}. From \cite{Facchinei07}, we know that if the sets $ \Psi_k(Q_{-k}) $ are defined by a system of inequalities, then there exists a function $\underline{h}(Q)$ which is component-wise convex with respect to $Q_k$ for all $k=1,2,...,K$ and furthermore
$ \mQ = \left \{ Q | h_j(Q) \geq 0, ~\forall j  \right \} $. In other words, jointly convex GNEPs are characterized by the fact that all the players have the same common constraints.

Rosen \cite{Rosen65} allows for a discriminatory treatment of players through the introduction of weights $r_i > 0, ~ i = 1,2,...,K$ with which the enforcer of the joint constraint can value each player's payoff. The main role of the weights in controlling the player's behavior is that they modify the Karush-Kuhn Tucker (KKT) multipliers of each player's utility maximization and entice the players to choose actions that lead to a desirable equilibrium outcome (among the infinitely many possible).

\subsection{Characterization of jointly convex GNEPs}
\label{subsec:GNEP_characterization}
We first define the normalized equilibria (NoE) of a jointly convex GNEP and then explore the relationship between GNEs and NoEs for this class of GNEPs. Consider the weighted utility function
\begin{equation}
f(B,Q,\underline{r}) = \sum_{i=1}^{K} r_i v_i(Q_1,...,Q_{i-1},B_i,Q_{i+1},...,Q_{K}),
\end{equation}
for a fixed vector of positive weights $\underline{r} = (r_1,r_2,...,r_K)$. The $K$-tuple $Q^{*} = (Q^{*}_1,\hdots,Q^{*}_K)$ is a normalized equilibrium (NoE) with respect to the weights $\underline{r}$, if $Q^{*}$ satisfies the \emph{fixed point} condition
\begin{equation}
Q^{*} = \arg \max_{B} f(B,Q^{*},\underline{r}),
\label{e:GNEP_fixed_pt}
\end{equation}
Let $h_j(Q_1,Q_2,...,Q_K) \geq 0$, $ j=1,...,J$ denote the set of joint constraints that each player must satisfy in addition to individual constraints on his control variables. Then, $Q^{*}$ is an NoE if and only if it satisfies the following KKT conditions: 
\begin{eqnarray}
&& r_k \nabla_k v_k(Q^{*}_k,Q^{*}_{-k}) + \sum_{j=1}^{J} \lambda^{*}_{kj} \nabla_k h_j(Q^{*}) = 0, \nonumber \\
&& \lambda^{*}_{kj} \geq 0, ~ h_j(Q^{*}) \geq 0,  ~\lambda^{*}_{kj} h_j(Q^{*}) = 0, ~ \forall ~ k,
\end{eqnarray}
where $\nabla_k$ is the derivative w.r.t $Q_k$ and $\lambda_{kj}$  are the Lagrange multipliers. The parameters $ \lambda_{kj} $ are called the \emph{shadow prices} of the joint constraints. For a general GNE, the Lagrange multipliers are unrelated to each other. The NoEs of the GNEP are a subset of GNEs for which the shadow prices of each of the joint constraints are equal for all players. 

\begin{mydef}
A given GNE of a GNEP is an NoE if and only if $ \lambda_{kj} = \lambda_{j}  $ for all $k=1,2,\hdots, K $, \textit{i.e.,} the shadow prices of each constraint are equal for each player.
\end{mydef}

For an NoE, the parameter $ \frac{\lambda_j}{r_i} $ is defined as the \emph{real price} for satisfying the $jth$ constraint by the $kth$ player. We next characterize the role of the weight vector $\underline{r}$. If the weight of the $kth$ player $r_k$ is greater than that of his competitors, then his real price is reduced and the marginal cost for constraint violation is lower than a player with higher weight. In other words, the choice of the vector $\underline{r}$enables the decision maker to decide how the burden of constraint satisfaction is to be divided among all the players in a GNEP.

\begin{prop}
The set of GNEs and NoEs is identical for a GNEP with only one jointly convex constraint.
\end{prop}
\begin{IEEEproof}
For a GNEP with only one jointly coupled constraint, the KKT conditions at a given GNE satisfy
\begin{eqnarray}
&& r_k \nabla_k v_k(Q^{*}_k,Q^{*}_{-k}) + \lambda^{*}_{k1} \nabla_k h_1(Q^{*}) = 0, \nonumber \\
&& \lambda^{*}_{k1} \geq 0, ~ h_j(Q^{*}) \geq 0,  ~\lambda^{*}_{k1} h_1(Q^{*}) = 0, ~ \forall ~ k.
\end{eqnarray}
We observe that this GNE can be considered an NoE for the above problem for the modified weight vector $ ( \frac{r_1}{\lambda_{11}}, \cdots,  \frac{r_K}{\lambda_{K1}} ) $ with unit Lagrange multipliers for the constraints. Thus we observe that the set of GNEs for this problem is a subset of NoEs. From the definition of NoEs, we know that the set of NoEs is a subset of the set of GNEs of a GNEP. This transformation between the GNEs and NoEs shows that the set of NoEs for a GNEP with one jointly convex constraint is identical to the set of GNEs. However, this can not be said about GNEPs with more than one jointly convex constraint.
\end{IEEEproof}

\textit{Note:} In this paper, we consider the MIMO BC and MIMO sum power MAC with only one joint sum power constraint and hence the set of NoEs and GNEs are identical for our problem. 

Once the decision maker of a jointly convex GNEP has established a desired NoE, the equilibrium implementation of the problem is as follows. The real prices associated with the joint constraints are used as penalty tax rates for constraint violation and the players have to allow for penalties in their payoffs. Define the penalty function:
\begin{equation}
T_k( \underline{\lambda}^{*},r_k,Q_1,...,Q_K) = \sum_{j=1}^{J} \frac{\lambda_j^{*}}{r_k}\mbox{max}(0,-h_j(Q_1,...,Q_K)),
\end{equation}
where $\lambda_j^{*}$ is the (equal for all players) Lagrange multiplier associated with the $jth$ common constraint, $r_k$ the weight of the $k^{th}$ player that defines the responsibility for constraint satisfaction. Clearly, a player with a large weight has a smaller penalty for violating the common constraint while a player with smaller weight suffers a larger penalty. If the constraint is perfectly satisfied, then the penalty is zero for all the players. Now, using the penalty function $T_k$, we define a game with modified utility functions which take into account the penalty for common constraint violation.
\begin{equation}
{\tilde v}_k(Q_k,Q_{-k}) = v_k(Q_k,Q_{-k}) - T_k( \underline{\lambda^{*}},r_k,Q_1,...,Q_K).
\end{equation}
Each player now needs to satisfy only his individual constraints and thus with the modified utility function, the problem becomes a traditional Nash equilibrium problem with \emph{decoupled constraints}. From \cite{Jacek09}, we know that the Nash equilibria for the modified utility functions are the same as the NoEs of the original problem with jointly convex constraints.

\subsection{Uniqueness of NoEs}
We first define several terms which help us in deriving a sufficient condition for the uniqueness of NoEs for a GNEP. Let
\begin{equation}
\sigma(Q,\underline{r}) = \sum_{i=1}^{K} r_i v_i(Q_i,Q_{-i}), ~ r_i > 0,
\end{equation}
to be the weighted sum of the utilities of each player, where $Q$ are the control variables for all the players and $\underline{r}$ is a vector containing a set of positive weights.
\begin{mydef} \cite{Rosen65}
The function
\begin{equation}
g(Q,\underline{r}) = \left [ \begin{array}{c} r_1 \nabla_1 v_1(Q_1,Q_{-1}) \\  r_2 \nabla_2 v_2(Q_2,Q_{-2}) \\ \vdots \\  r_K \nabla_K v_K(Q_k,Q_{-k}) \end{array}
           \right ],
\end{equation}
where $\nabla_i$ is the derivative w.r.t the $i^{th}$ players'control variables is called the \emph{pseudo-gradient} of $\sigma(Q,r)$.
\end{mydef}

\begin{mydef} (defined for scalar variables in \cite{Rosen65})
The function $\sigma(Q,\underline{r})$ is called \emph{diagonally strictly concave} (DSC) in matrix valued strategies for a fixed $ \underline{r} $ > 0, if for every $ \tQ, \hQ \in \mQ $, we have that
\begin{equation}
\mbox{Tr} \left [ (\tQ - \hQ)^{T} g(\hQ,\underline{r}) + (\hQ - \tQ)^{T} g(\tQ,\underline{r}) \right ] > 0.
\end{equation}
\label{def:DSC}
\end{mydef}

\begin{prop}
If the utility functions of a jointly convex GNEP satisfy the DSC condition for any given $\underline{r}$, then the GNEP has a unique NoE for that given value of $\underline{r}$.
\label{prop:DSC_uniqueness}
\end{prop}
The proof of this proposition depends on the nature of the jointly convex constraints, in general. To simplify the presentation and illustrate the key ideas in the proof, we consider a constraint of the form $\sum_{i=1}^{K} \mbox{Tr} [Q_i] \leq P_{tot}$. We note that this proof technique can be directly applied to problems with multiple and heterogeneous constraints.
\begin{IEEEproof}
We assume that there exist multiple NoEs for the given weight vector $\underline{r}$ and then arrive at a contradition to show the DSC property ensures uniqueness of NoEs. Let us assume that $ \tQ =  ( {\tilde Q}_1,{\tilde Q}_2,...,{\tilde Q}_K ) $ and $\hQ = ( {\hat Q}_1,{\hat Q}_2,...,{\hat Q}_K ) $ be two $K$-tuples of covariance matrices which are NoEs to the game characterized by the weight vector $\ur$. We know from (\ref{e:GNEP_fixed_pt}) that
\begin{equation}
\tQ = \arg \max_{B} f(B,\tQ,\underline{r}) ~  \mbox{and} ~ \hQ = \arg \max_{B} f(B,\hQ,\underline{r}) .
\end{equation}
Writing the Lagrangian for the maximization of the weighted utility function, we get
\begin{equation}
\mathcal{L} = f(B,Q,\ur) + \lambda ( \sum_{i=1}^{K} \mbox{Tr} [Q_i] - P_{tot} )  + \sum_{i=1}^{K} \mbox{Tr} [L_i Q_i ]
\end{equation}
 Writing the Karush-Kuhn-Tucker (KKT) conditions \cite{Boyd} for the two equilibria yields:
\begin{enumerate}[(a)]
\item $\tQ_i,~ \hQ_i \succeq 0, ~ i=1,2,...,K$
\item $\sum_{i=1}^{K} \mbox{Tr} [ \tQ_i ] \leq P_{tot} $ and $\sum_{i=1}^{K} \mbox{Tr} [ \hQ_i ] \leq P_{tot} $.
\item $\mbox{Tr} [ \tL_i \tQ_i ] = 0$ and $\mbox{Tr} [ \hL_i \hQ_i ] = 0$.
\item $ {\tilde \lambda} \left ( \sum_{i=1}^{K} \mbox{Tr} [\tQ_i ] - P_{tot}  \right) = 0$.
\item $ {\hat \lambda}   \left ( \sum_{i=1}^{K} \mbox{Tr} [\hQ_i ] - P_{tot}  \right) = 0$.
\item $ r_i \nabla_i v_i(\tQ) + \tL_i - \tl I = 0$
\item $ r_i \nabla_i v_i(\hQ) + \hL_i - \hl I = 0$.
\end{enumerate}
where $ \tl, \hl \geq 0 $ and $ \tL_i, \hL_i \succeq 0 $ are the Lagrange multipliers associated with sum power constraint and the positive semi-definiteness of the solutions respectively. Now multiplying (f) and (g) with $ (\hQ_i - \tQ_i) $ and $ (\tQ_i - \hQ_i) $ respectively, summing on $i$ and taking the trace we get
\begin{align}
0 & = \sum_{i=1}^{K} \mbox{Tr} \left [ (\hQ_i - \tQ_i)(  r_i \nabla_i v_i(\tQ) + \tL_i - \tl I ) \right ] \nonumber \\
  & ~~~~ + \sum_{i=1}^{K} \mbox{Tr} \left [ (\tQ_i - \hQ_i)(  r_i \nabla_i v_i(\hQ) + \hL_i - \hl I ) \right ] \nonumber \\
	\end{align}
\begin{align}	
 0 & = \sum_{i=1}^{K} \mbox{Tr} \left [ (\hQ_i - \tQ_i) r_i \nabla_i v_i(\tQ) + (\tQ_i - \hQ_i) r_i \nabla_i v_i(\hQ) \right ] \nonumber \\
  & ~~~~ + \sum_{i=1}^{K} \mbox{Tr} \left [ (\hQ_i - \tQ_i) ( \tL_i - \tl I )  +  (\tQ_i - \hQ_i)( \hL_i - \hl I ) \right ] \nonumber \\
  & = \alpha + \beta.
\end{align}
Re-arranging and evaluating the second term,
\begin{align}
& \beta = \textrm{Tr} \left [ \sum_{i=1}^{K} ( {\tQ}_i - {\hQ}_i ) \left \{ (\tl I - \tL_i) - (\hl I - \hL_i)  \right \}  \right ] \nonumber \\
& ~ \stackrel{(c)}{=} \textrm{Tr} \left [ \sum_{i=1}^{K} ( \tl \tQ_i  - \hl \tQ_i  + \tQ_i \hL_i - \tl \hQ_i  + \hQ_i \tL_i + \hl \hQ_i ) \right ] \nonumber \\
& ~\stackrel{(d,e)}{=} \tl P_{tot} + \hl P_{tot} - \textrm{Tr} \left [ \hl \sum_{i} \tQ_i  +  \tl \sum_{i}  \hQ_i   \right ] \nonumber \\
&	~~~ + \textrm{Tr} \left [ \sum_{i} (\tQ_i \hL_i + \hQ_i \tL_i)  \right ] \nonumber \\
& ~ \stackrel{(a)}{\geq} \textrm{Tr} \left [ \hl \left (P_{tot} - \sum_{i} \tQ_i \right )  \right ] + \textrm{Tr} \left [ \tl \left (P_{tot} - \sum_{i} \hQ_i \right ) \right ] \nonumber \\
& ~ \stackrel{(b)}{\geq} 0.
\end{align}
We have shown that $ \beta \geq 0 $ and hence for $ \alpha + \beta = 0 $ we need that $ \alpha \leq 0 $. Now
\begin{align}
\alpha & = \sum_{i=1}^{K} \mbox{Tr} \left [ (\hQ_i - \tQ_i) r_i \nabla_i v_i(\tQ) + (\tQ_i - \hQ_i) r_i \nabla_i v_i(\hQ) \right ] \nonumber \\
       & = \mbox{Tr} \left [ (\hQ - \tQ)^{T} g(\tQ,\underline{r}) + (\tQ_i - \hQ_i)g(\hQ,\underline{r}) \right ].
\end{align}
Recognizing that $ \alpha > 0 $ is the DSC condition from Definition \ref{def:DSC}, we arrive at a contradiction as we assume that the DSC condition is satisfied for the given value of $\ur$. Hence, when the DSC condition is satisfied, the NoE of the GNEP for a given weight vector $\ur$ is unique.
\end{IEEEproof}
Finally, we observe that the DSC condition is one of the sufficient conditions for determining the uniqueness of NoEs and there could exist several other sufficient conditions.

%% file: game_model.tex
Consider a multiple-input multiple-output (MIMO) BC with $K$ receivers. The transmitted signal, denoted by $\underline{x}_{n_t \times 1}$, where $n_t$ is the number of TX-antennas, is the sum of independent $\underline{x_i}$, each drawn from a Gaussian codebook and intended for the $i^{th}$ receiver (RX): $ \underline{x} = \sum_{i=1}^{K} \underline{x}_i, ~ \underline{x}_i \sim \mathcal{N}(0,Q_i)$. The received signal at the $i^{th}$ RX can be expressed as
\begin{equation}
\underline{y}_i = \acute{H}_i \underline{x} + \underline{n}_i, ~ \underline{z}_i \sim \mathcal{N}(0,N_i),
\end{equation}
where ${\acute{H}_i} $ is the ${n_i \times n_t}$ channel gain matrix from the TX to the $i^{th}$ RX and $n_i$ is the number of antennas at the $i^{th}$ RX. Note that this representation is not unique and the received signal can also be expressed as 
\begin{equation}
\underline{y}_i = H_i \underline{x} + \underline{n}_i, ~ \underline{z}_i \sim \mathcal{N}(0,N_0I),
\end{equation}
and $ \acute{H}_i = \sqrt{N_0} N^{-1/2}_i H_i $. Both these representations are used in the paper and the choice of representation will be indicated based on the context. The signal at each RX is a linear transformation of the sum of the signals intended for all the receivers.  Without loss of generality, we assume that TX signaling is constrained by a sum power constraint
$$ \mbox{Tr} \left( E[ \underline{x} \underline{x}^{T} ] \right) = \sum_{i=1}^{K} \mbox{Tr} \left [  Q_i \right ]  \leq P_{tot}, $$
where $P_{tot}$ is the maximum transmit sum power for all the antennas. We begin by assuming that the transmitter performs dirty paper coding with a fixed encoding order $\underline{\pi} = \{K,K-1,...,1\}$. The rate achieved by the $k^{th}$ receiver, which is considered as a player in the BC game, can be expressed as
\begin{equation}
v_k(Q_k,Q_{-k}) = \mbox{log} \left ( \frac{| N_k + \acute{H}_k (\sum_{i=1}^{k} Q_k ) \acute{H}^{H}_k |}{ |  N_k + \acute{H}_k (\sum_{i=1}^{k-1} Q_k ) \acute{H}^{H}_k |  } \right ),
\label{e:BC_util_1}
\end{equation}
or equivalently
\begin{equation}
v_k(Q_k,Q_{-k}) = \mbox{log} \left ( \frac{| N_0 I + H_k (\sum_{i=1}^{k} Q_k ) H^{H}_k |}{ |  N_0 I + H_k (\sum_{i=1}^{k-1} Q_k ) H^{H}_k |  }
\right ).
\label{e:BC_util_2}
\end{equation}
In this paper, the utility of each player is defined as the achievable rate under the BC with dirty paper coding and a given encoding order. Note that $v_k(Q_k,Q_{-k})$ is concave in $Q_k$ and continuous in $Q_i$ for all $i=1,...,K$. The controlling variables for the players $Q_k$ are coupled via a sum power constraint which is common to every receiver and hence the BC game is a GNEP with jointly convex constraints (see Definition \ref{def:jointlyconvex}). 

\subsubsection{Uniqueness of NoEs}
We now examine the existence and uniqueness of NoEs for the MIMO BC (As noted in Section \ref{sec:intro}, the set of NoEs and GNEs for the problems considered in this paper are identical and both terms are used interchangingly to match the usage in related literature). Uniqueness of NoEs is a desirable property of a game as it ensures that every player can accurately predict the outcome of the game and there is no ambiguity about choosing from a set of equilibria. From Theorem $3$ of Rosen \cite{Rosen65}, we know that there exists a NoE for a concave $K$-person game for every weight vector $\underline{r} > 0$. As the BC game is a concave game with jointly convex constraints, we immediately conclude that there exists an NoE for every weight vector $\underline{r} > 0$. To analyze the uniqueness of NoEs, we begin by considering a special class of BCs called  \emph{aligned and degraded broadcast channels} (ADBCs) \cite{Weingarten06}.

\begin{mydef}
A MIMO BC is \textit{aligned} and \textit{degraded} if the BC is aligned, \textit{i.e.,} $ n_t = n_1 = n_2 = ... = n_K $ and $\acute{H}_i = I_{n_t \times n_t} $ and the covariances of the Gaussian noise at the receiver are ordered such that $ 0 \prec N_1 \preceq N_2 \preceq ... \preceq N_K $, where $ A \preceq B $ implies that $B-A$ is a positive semi-definite matrix \cite{Weingarten06}.
\end{mydef}
For an ADBC, the achievable rate or the utility function of the $k^{th}$ player (from (\ref{e:BC_util_1})) simplifies to
\begin{equation}
v^{ADBC}_k(Q_k,Q_{-k}) = \mbox{log} \left ( \frac{| N_k + \sum_{i=1}^{k} Q_i |}{| N_k + \sum_{i=1}^{k-1} Q_i  |} \right ).
\label{e:util_DPC_ADBC}
\end{equation}

We know from Section \ref{sec:preliminaries} that the DSC criterion in Definition (\ref{def:DSC}) is a sufficient condition for the uniqueness of NoE for a given positive weight vector $\ur$. We now determine the weight vectors $\ur$ and the encoding orders $\underline{\pi}$ for which the DSC criterion holds for the ADBC. We first state two trace inequalities that will be used to derive the uniqueness results.

\begin{lemma} \cite{Belmega11ineq}
For any positive integer $K$ and a set of positive semi-definite matrices $A_1,A_2,...,A_K$ and $B_1,B_2,...,B_K$ such that $A_1 \succ 0$ and $B_1 \succ 0$, we have that
\begin{align}
\textrm{Tr} \left \{ \sum_{k=1}^{K} (A_k - B_k) \left [ \left (\sum_{l=1}^{k} B_l  \right )^{-1} - \left (\sum_{l=1}^{k} A_l \right )^{-1}  \right ]  \right \} \geq 0.
\label{e:traceineq}
\end{align}
\label{lemma:matrix_ineq_1}
\end{lemma}
Note that the above set of inequalities is not tight in general and it is possible to derive tighter inequalities in some scenarios. For example, for $K = 2$ and any positive real number $w$, it has been shown in \cite{Furuichi11ineq} that
\begin{align}
\mbox{Tr} &  [ (A_1 - B_1) ( B_1^{-1}  - A_1^{-1} )  \nonumber \\
     & + 4 (A_2 - B_2)\left \{ (wB_1 + B_2)^{-1} - (wA_1 + A_2)^{-1}  \right \} ] \geq 0.
\label{e:traceineq_2}
\end{align}

\begin{thm}
For a $K$-receiver ADBC with dirty paper coding at the TX and encoding order $\underline{\pi} = \{K,K-1,...,1\}$, a unique \emph{NoE achieving strategy} exists for every weight vector $ \underline{r} $ which satisfies $r_1 \geq r_2 \geq ... \geq r_K > 0$.
\label{thm:unique_K_ADBC}
\end{thm}
\begin{IEEEproof}
Let $(\tQ_1,\tQ_2,...,\tQ_k)$ and $(\hQ_1,\hQ_2,...,\hQ_k)$ be any two distinct vectors of covariance matrices which satisfy the sum power constraint: $ \sum_{i=1}^{K} \mbox{Tr} [ \tQ_i ] \leq P_{tot} $ and $ \sum_{i=1}^{K} \mbox{Tr} [\hQ_i ] \leq P_{tot} $. Substituting the utility function for the ADBC from (\ref{e:util_DPC_ADBC}) in the DSC condition in Definition \ref{def:DSC}, we get
\begin{align}
& \textrm{Tr} \left [ \sum_{k=1}^{K} r_k ( {\hQ}_k - {\tQ}_k ) \left \{  \nabla_k v_k({\tQ}) - \nabla_k v_k({\hQ})   \right \} \right ] \nonumber \\
& = \textrm{Tr} \Bigg [ \sum_{k=1}^{K} r_k ( {\hQ}_k - {\tQ}_k ) \bigg \{ (N_k + \sum_{i=1}^{k} \tQ_i  )^{-1}  \nonumber \\
& ~~~~~~~~~~~~~~~~~~~~~~~~~~~~~~~~~~~~~~~~ - (N_k + \sum_{i=1}^{k} \hQ_i  )^{-1} \bigg \} \Bigg ] \nonumber \\
& = \sum_{n=1}^{K-1} (r_{n} - r_{n+1}) \mT_n + r_K \mT_{K},
\end{align}
where the term $\mT_n$ can be expressed as
\begin{equation}
\textrm{Tr} \Bigg [  \sum_{k=1}^{n} (\hQ_k - \tQ_k) \bigg \{ (N_k + \sum_{i=1}^{k} \tQ_i)^{-1} - (N_k + \sum_{i=1}^{k} \hQ_i)^{-1} \bigg \} \Bigg ].
\end{equation}	
Thus proving that $\mT_n > 0 $ is sufficient to determine the uniqueness of NoE for every $ r_n \geq r_{n+1} $. Notice that the structure of $\mT_n$ closely resembles the inequality in (\ref{e:traceineq}). We choose the quantities $A_1 = N_1 + \tQ_1$, $B_1 = N_1 + \hQ_1$, $A_i = N_i - N_{i-1} + \tQ_i$ and $B_i = N_i - N_{i-1} + \hQ_i$. By definition, since $N_1$ is a positive definite matrix and $\tQ_1$, $\hQ_1$ are positive semi-definite the matrices $A_1$ and $B_1$ are strictly positive definite. From the degradedness of the channel, we get that $N_i - N_{i-1}$ is a positive semi-definite matrix and hence $A_i$ and $B_i$ are positive semi-definite for $i=2,...,K$. Substituting the values of $A_i$ and $B_i$ in (\ref{e:traceineq}), it is straight forward to see that $\mT_n \geq 0$. For an ADBC channel having identity channel matrices, we know from \cite{Belmega10} that if the $\tQ \neq \hQ$, then $ \mT_n > 0 $ and hence the NoEs of the ADBC game are unique for $r_1 \geq r_2 ... \geq r_K > 0 $ and encoding order $\underline{\pi}=\{K,K-1,...,1\}$.
\end{IEEEproof}
It is clear that the weight vectors $\ur$ for which uniqueness can be shown are dependent on the tightness of the matrix trace inequalities. For $K=2$, we know from the literature that the inequality in (\ref{e:traceineq}) has been generalized to the inequality in (\ref{e:traceineq_2}). Using this tighter inequality, we now show that uniqueness of NoEs for the $2$-user ADBC can be determined for a larger set of vectors. In addition, note that the uniqueness of NoEs is true only for the specified decoding order and the weight vectors and the uniqueness of NoEs cannot be guaranteed for any other decoding order and weight vectors.

\begin{thm}
For a $2$-user ADBC with dirty paper coding at the transmitter and interference canceling receivers, a unique NoE exists for $r_1 \geq r_2/4 > 0$.
\end{thm}
\begin{IEEEproof}
The proof follows exactly on the lines of Theorem \ref{thm:unique_K_ADBC} with the DSC condition decomposing into two terms given by $(r_1 - \frac{r_2}{4}) T_1$ and $\frac{r_2}{4} T_2$. Now using the inequality in (\ref{e:traceineq_2}) with $w=1$, it is easy to show that there exists a unique NoE for each weight vector which satisfies $r_1 \geq r_2/4 > 0$.
\end{IEEEproof}

We now consider the uniqueness of NoEs for the general Gaussian BC. Due to the matrix valued nature of the problem and the possibility of a null space existing for the channel gain matrices $H_i$, we make a distinction between the uniqueness of \emph{NoE achieving strategies} and the uniqueness of \emph{NoE utility} for the general Gaussian BC. While the former always implies the later, it is not always true that the uniqueness of NoE utilities implies the uniqueness of NoE acheiving strategies. 
From the derivation of the DSC condition in Proposition \ref{prop:DSC_uniqueness}, we know that if there exist at least two NoEs, $\tQ = (\tQ_1,\tQ_2,...,\tQ_K) $ and $\hQ = (\hQ_1,\hQ_2,...,\hQ_K) $ for a game with jointly convex coupled constraints and a given weight vector $\ur$, then
\begin{align}
&\sum_{k=1}^{K} r_k \mbox{Tr} \left [ (\hQ_k - \tQ_k) \nabla_k v_k(\tQ_k)  + (\tQ_k - \hQ_k) \nabla_k v_k(\hQ_k) \right ] \nonumber \\
& = \alpha \leq 0.
\label{e:GBC_uniqueness}
\end{align}

\begin{prop}
For a $K$-receiver BC with dirty paper coding at the TX and encoding order $\underline{\pi} = \{\pi_1,\pi_2,...,\pi_K\}$, a weight vector $ \ur$ such that  $ r_{\pi_K} \geq r_{\pi_{K-1}} ... \geq r_{\pi_{1}} > 0 $ and any two feasible strategies $\tQ = (\tQ_1,\tQ_2,...,\tQ_K) $ and $\hQ = (\hQ_1,\hQ_2,...,\hQ_K) $, we have that
\begin{align}
&\sum_{k=1}^{K}  r_k \mbox{Tr} \left [  (\hQ_k - \tQ_k) \nabla_k v_k(\tQ_k)  + (\tQ_k - \hQ_k) \nabla_k v_k(\hQ_k)  \right ] \nonumber \\
& = \alpha \geq 0.
\label{e:BC_condition}
\end{align}
\label{prop:BC_uniq}
\end{prop}
\begin{IEEEproof}
Without loss of generality, we assume that $\underline{\pi} = \{K,K-1,...,1\}$. The proof for all the other decoding orders can be directly obtained by following the procedure shown for the given decoding order. Consider two $K$-tuples of covariance matrices $\tQ = (\tQ_1,\tQ_2,...,\tQ_K) $ and $\hQ = (\hQ_1,\hQ_2,...,\hQ_K) $ satisfying the sum power constraint. Define the matrix valued set of functions $\phi_k(Q) = ( H_k  (\sum_{i=1}^{k} Q_i) H_k^{H} + N_0 I)^{-1}, ~ k=1,...,K$. By re-arranging terms and substituting the utility function for the BC from (\ref{e:BC_util_2}) in $\alpha$, we get
\begin{align}
& \sum_{k=1}^{K} r_k \textrm{Tr} \left [  ( {\hQ}_k - {\tQ}_k ) \left (  \nabla_k v_k({\tQ}) - \nabla_k v_k({\hQ})   \right ) \right ] \nonumber \\
& = \sum_{k=1}^{K} r_k \textrm{Tr} \left [  ( {\hQ}_k - {\tQ}_k ) \left ( H^{H}_k \phi_k(\tQ) H_k  -  H^{H}_k \phi_k(\hQ) H_k  \right ) \right ] \nonumber \\
& = \sum_{k=1}^{K} r_k \textrm{Tr} \left [  ( H_k ({\hQ}_k - {\tQ}_k) H^{H}_k )( \phi_k(\tQ) - \phi_k(\hQ) ) \right ] \nonumber \\
& = \sum_{n=1}^{K-1} (r_{n} - r_{{n+1}}) \mT_n + r_K \mT_{K},
\end{align}
where $\mT_n = \textrm{Tr} \left [  \sum_{k=1}^{n} H_k (\hQ_k - \tQ_k) H^{H}_k (\phi_k(\tQ) - \phi_k(\hQ)) \right ] $. Choose the quantities $A_1 = N_0 I + H_1 \tQ_1 H^{H}_i $, $B_1 = N_0 I + H_1 \hQ_1 H^{H}_1$, $A_i = H_i \tQ_i H^{H}_i $ and $B_i = H_i \hQ_i H^{H}_i $. By definition, as $N_0 I$ is a positive definite matrix and $\tQ_1$, $\hQ_1$ are positive semi-definite, the matrices $A_1$ and $B_1$ are strictly positive definite. In addition, we see that $A_i $ and $B_i$ are positive semi-definite matrices for $i=2,...,K$. Using the trace inequality in Lemma \ref{lemma:matrix_ineq_1} which states that $ \sum_{i=1}^{N} \mbox{Tr} \Big [ (A_i - B_i) \times \Big \{ (\sum_{j=1}^{i} B_j)^{-1} - (\sum_{j=1}^{i} A_j)^{-1} \Big \} \Big ] \geq 0 $, with equality only when $A_i = B_i, ~ \forall i = 1,2,...,N$, we see that $\mT_n \geq 0 $ for all $n$, thus proving the condition in (\ref{e:BC_condition}).
\end{IEEEproof}

\begin{thm}
The NoE utility of the general Gaussian BC with a encoding order $\underline{\pi} = \{\pi_{K},\pi_{K-1},...,\pi_{1}\} $ is unique for all weight vectors $\ur$ such that $r_{\pi_K} \geq r_{\pi_{K-1}} ... \geq r_{\pi_1} > 0 $. However, the NoE achieving strategies (strategy is choice of $Q = (Q_1,Q_2,...,Q_k)$) are not unique in general.
\label{thm:unique_K_BC}
\end{thm}
\begin{IEEEproof}
From (\ref{e:GBC_uniqueness}) and (\ref{e:BC_condition}), we infer that if $\tQ = (\tQ_1,\tQ_2,...,\tQ_K) $ and $\hQ = (\hQ_1,\hQ_2,...,\hQ_K) $ are two NoEs (and hence are also achievable strategies), then $ \alpha = 0$. Now from Proposition \ref{prop:BC_uniq} we know that
\begin{equation}
\alpha = 0 \Leftrightarrow A_i = B_i ~ \forall i \Leftrightarrow H_i \tQ_i H^{H}_i = H_i \tQ_i H^{H}_i.
\end{equation}
Substituting in the utility function in (\ref{e:BC_util_2}), we observe that given a weight vector $\underline{r}$ such that $r_{\pi_K} \geq r_{\pi_{K-1}} \geq ... \geq r_{\pi_1} > 0$ and encoding order $ \pi = \{\pi_K,\pi_{K-1},...,\pi_1]$, the NoE utility obtained by both the NoE achieving strategies $\tQ$ and $\hQ$ is identical implying the \emph{uniqueness of the NoE utility} for a weight vector and its associated encoding order. 

If $H_i$ is a square matrix with full rank, it can be easily shown that $H_i \tQ_i H^{H}_i = H_i \tQ_i H^{H}_i$ implies that $ \tQ = \hQ $. In other words, for a full rank invertible square channel matrix, the uniqueness of NoE utility also implies the uniqueness of NoE strategy. However, this is not true for all channel matrices and hence the NoE achieving strategies for the BC are not unique in general.
\end{IEEEproof}

To summarize, we have determined the combinations of encoding orders and weight vectors for which the \emph{NoE utility} for the Gaussian BC is unique. For all other scenarios, the uniqueness of NoE utility cannot be guaranteed as the DSC condition from Proposition \ref{prop:DSC_uniqueness} is only a sufficient condition for determining the uniqueness of NoEs. For the special case of an ADBC, we can show that for the encoding order which achieves the capacity region and a subset of weight vectors ordered accordingly, the NoE achieving strategy is unique.

\section{The Sum Power MAC as a Generalized Nash Equilibrium Problem}
\label{sec:sumpowerMAC}
The sum power MAC and the BC are closely related problems in information theory by the duality result derived in \cite{Vishwanath03}. In particular, it is shown in \cite{Vishwanath03} that the capacity region of the general Gaussian BC is identical to the capacity region of a suitably defined MIMO MAC with a joint sum power constraint across all TXs, in contrast to the individual power constraints for each TX for a regular MAC. In this section, we characterize the NoEs of the sum power MAC, which will enable us to derive a game-theoretic duality between the MAC and the BC in Section \ref{sec:duality} of this paper.  

\subsection{Signal Model and the Sum Power MAC game}
Consider a MIMO MAC channel with $K$ TXs (users) with the $i^{th}$ TX has $n_i$ antennas and a common receiver with $n_r$ antennas. Each transmitted signal is drawn from a Gaussian codebook $\underline{x}_i \sim \mN(0,Q_i)$ and transmitted signals satisfy a sum power constraint given by $  \sum_{i=1}^{K} \mbox{Tr} \left [ Q_i \right ] \leq P_{tot} $. The received signal $\underline{y}$ can be written as
\begin{equation}
\underline{y} = \sum_{i=1}^{K} H_i \underline{x}_i + \underline{z},
\end{equation}
where $H_i$ is $ n_r \times n_i$ channel matrix from the $i^{th}$ TX to the RX and $\underline{z} \sim \mN(0,N_0I) $ is the AWGN at the receiver. Each TX is considered as a player in the game with an action space that consists of signaling covariance matrices which satisfy the joint sum power constraint and hence the sum power MAC game can be modeled as a GNEP. The receiver performs successive interference cancellation to decode the signals from each TX and without loss of generality, we assume that $\pi = [K,K-1,...,1] $ is the decoding order.
The utility function for each TX is the rate it obtains and can be expressed as
\begin{equation}
v^{MAC}_k(Q_k,Q_{-k}) = \mbox{log} \left (  \frac{| N_0 I + \sum_{i=1}^{k} H_i Q_i H^{H}_i |}{|  N_0 I + \sum_{i=1}^{k-1} H_i Q_i H^{H}_i|} \right )
\label{e:SPMAC_util}
\end{equation}
Clearly, the utility function for the $k^{th}$ player (TX) is concave in $Q_k$ and a continuous function of all other variables. In addition, as the joint sum power constraint defines a convex set of feasible strategies, the sum power MAC is a concave game and hence from \cite{Rosen65} we know that for every weight vector $\underline{r}$, there exists at least one NoE achieving strategy.

\subsection{Uniqueness of NoEs}
In Section \ref{sec:game_model}, we determined the uniqueness of NoE acheiving strategies and utilities for the ADBC and general Gaussian BCs. Using similar techniques, we now examine the uniqueness of NoEs for the sum power MAC. As in the case of the BC, we make a distinction between the uniqueness of \emph{NoE acheiving strategies} and the uniqueness of \emph{NoE utilities} for the sum power MAC. 

\begin{prop}
For the sum power MAC with a successive interference canceling receiver implementing a decoding order of $\underline{\pi} = \{\pi_1,\pi_{2},...,\pi_K\}$, a weight vector $\underline{r}$ such that $r_{\pi_K} \geq r_{\pi_{K-1}} \geq ... \geq r_{pi_1} > 0$ and any two feasible strategies $\tQ = (\tQ_1,\tQ_2,...,\tQ_K) $ and $\hQ = (\hQ_1,\hQ_2,...,\hQ_K) $, we have that
\begin{align}
&\sum_{k=1}^{K}  r_k \mbox{Tr} \left [  (\hQ_k - \tQ_k) \nabla_k v_k(\tQ_k)  + (\tQ_k - \hQ_k) \nabla_k v_k(\hQ_k)  \right ] \nonumber \\
& = \alpha \geq 0.
\label{e:SPMAC_2}
\end{align}
\label{prop:SPMAC}
\end{prop}
\begin{IEEEproof}
The proof of this Proposition follows on the same lines as the proof of Proposition \ref{prop:BC_uniq} with $\phi_k(Q) = ( \sum_{i=1}^{k} H_i Q_i H_i^{H} + N_0 I)^{-1} $ and is omitted for brevity. 
\end{IEEEproof}

\begin{thm}
The NoE utility of the sum power MAC with a decoding order $\underline{\pi} = \{\pi_1,\pi_2,...,\pi_K \} $ is unique for all weight vectors $\ur$ such that $r_{\pi_K} \geq r_{\pi_{K-1}} ... \geq r_{\pi_1} > 0 $. However, the NoE achieving strategies (strategy is choice of $Q = (Q_1,Q_2,...,Q_k)$) are not unique in general.
\end{thm}
\begin{IEEEproof}
We know from the DSC condition in Proposition \ref{prop:DSC_uniqueness} that if there exist two NoEs, $\tQ = (\tQ_1,\tQ_2,...,\tQ_K) $ and $\hQ = (\hQ_1,\hQ_2,...,\hQ_K) $ for a game with jointly convex coupled constraints and a given weight vector $\ur$, then
\begin{align}
&\sum_{k=1}^{K} r_k \mbox{Tr} \left [ (\hQ_k - \tQ_k) \nabla_k v_k(\tQ_k)  + (\tQ_k - \hQ_k) \nabla_k v_k(\hQ_k) \right ] \nonumber \\
& = \alpha \leq 0.
\end{align}
Now from (\ref{prop:SPMAC}) and the above condition, we infer that if $\tQ = (\tQ_1,\tQ_2,...,\tQ_K) $ and $\hQ = (\hQ_1,\hQ_2,...,\hQ_K) $ are two NoEs for a given weight vector $\ur$ (and hence are also achievable strategies), then $ \alpha = 0$. Now from Proposition \ref{prop:SPMAC} we know that
\begin{equation}
\alpha = 0 \Leftrightarrow A_i = B_i ~ \forall i \Leftrightarrow H_i \tQ_i H^{H}_i = H_i \tQ_i H^{H}_i.
\end{equation}
Substituting in the utility function in (\ref{e:SPMAC_util}), we observe that given a decoding order $\underline{\pi} = \{\pi_1,\pi_2,...,\pi_K \} $ and  weight vector $\underline{r}$ such that $r_{\pi_K} \geq r_{\pi_{K-1}} ... \geq r_{\pi_1} > 0 $, the NoE utility obtained by both the NoE achieving strategies $\tQ$ and $\hQ$ is identical implying the \emph{uniqueness of the NoE utility} for a weight vector and its associated decoding order. 

Similar to the general Gaussian BC, we observe that if $H_i$ is a square matrix with full rank, it can be easily shown that $H_i \tQ_i H^{H}_i = H_i \tQ_i H^{H}_i$ implies that $ \tQ = \hQ $. In other words, for a full rank invertible square channel matrix, the uniqueness of NoE utility also implies the uniqueness of NoE achieving strategies. However, this is not true for all channel matrices and hence the NoE achieving strategies for the sum power MAC are not unique in general.
\end{IEEEproof}

%% file: relationship.tex
In this section, we explore the relationship between the points on the Pareto-optimal boundary of the rate region of the BC and sum power MAC and the NoEs of the corresponding games. This is a necessary precursor towards establishing our duality result. In specific, we are interested in Pareto-efficient NoEs, \textit{i.e.,} equilibria which also maximize a weighted sum of utilities. \emph{If a rate point of the BC or sum power MAC is both Pareto-efficient and an NoE, then it is a socially optimum solution with self-enforcing properties and enables the implementation of optimal centralized solutions using distriubted and decentralized algorithms.} 

To determine Pareto-efficient NoEs, we first derive a relation between the weights $\underline{\gamma}$ which characterize a given Pareto-optimal solution and the weights $\underline{r}$ for which this rate point is a NoE utility \cite{Jacek09}. We generalize the procedure in \cite{Jacek09} to $K$-players and matrix valued strategies to derive the desired relationship.

\subsection{Pareto-Efficiency First Order Conditions}
Consider a regulator (for example, a base station or a coordinating entity) who would like to control the rates achievable by each user. The regulator would like choose a feasible strategy to optimize a weighted sum rate of utility functions. Let $\mathcal{F}$ denote the set of all K-tuples of covariance matrices which satisfy the sum power constraint. The regulator's problem can be written as
\begin{equation}
\max_{Q \in \mathcal{F}} \sum_{i=1}^{K} \gamma_i v_i(Q_i,Q_{-i}),
\end{equation}
where $v_i$ is the utility function and $\gamma_i$ is the weight attached to the $i^{th}$ player. We assume that the regulator is interested in solutions which saturate the constraint  $ \sum_{i=1}^{K} \mbox{Tr} [ Q_i ] \leq P_{tot} $. The Lagrangian for the regulator w.r.t the common constraint can be written as
\begin{equation}
L^{P} = \sum_{i=1}^{K} \gamma_i v_i(Q_i,Q_{-i}) - \mu \left [ \sum_{i=1}^{K} \mbox{Tr} [Q_i] - P_{tot}  \right ] + \sum_{i=1}^{K} \mbox{Tr}\left [L_i Q_i \right ],
\label{e:Pareto}
\end{equation}
where $\mu > 0 $ and $L_i \succeq 0$ are the dual variables for the constraints. As the feasible set $\mathcal{F}$ is non-empty for $P_{tot} > 0 $, the KKT conditions can be written for all $k=1,2,...,K$ as:
\begin{align}
& \frac{\partial L^{P}}{\partial Q_k} = \sum_{i=1}^{K} \gamma_i \frac{\partial v_i(Q_i,Q_{-i})}{\partial Q_k} - \mu I + L_k = 0 \nonumber \\
& L_k \succeq 0, ~ \mbox{Tr} [ L_k Q_k ] = 0 \nonumber \\
& \mu > 0, ~  \sum_{i=1}^{K} \mbox{Tr}\left [  Q_i \right ] - P_{tot}  = 0.
\label{e:Pareto_KKT}
\end{align}
Let us denote the Pareto-optimal solution to be $Q^{*}(\underline{\gamma}) = (Q_1(\underline{\gamma}),...,Q_K(\underline{\gamma}))$ where $\underline{\gamma} = (\gamma_1,...,\gamma_K)^T$. Given the concavity of the utility functions in the users' control variable and convexity of the constraint set, the above conditions are sufficient for the solution of the differential equations to be Pareto-optimal.

\subsection{Rosen's equilibrium first order conditions}
We know from Section \ref{subsec:GNEP_characterization} that it is possible to control players, who share a common constraint, to satisfy this constraint by interpreting the GNEP as a modified Nash equilibrium problem with penalties for violating the common constraint.  Mathematically, the regulator seeks a solution which is the NE of the modified game. The Lagrangian of the optimization associated with each player can be written as
\begin{equation}
L^{R}_k = r_k v_k(Q_k,Q_{-k}) - \lambda  \left [ \sum_{i=1}^{K} \mbox{Tr}[Q_i] - P_{tot} \right ] + \mbox{Tr} \left [ M_k Q_k \right ],
\label{e:rosen_Lagrangian}
\end{equation}
where $\lambda$ and $M_k$ are the dual variables for the constraints. Note that the factor $\lambda$ is equal for all players as we are interested in NoE where the shadow price ($\frac{\lambda}{r_i}$ is the true price, see Section \ref{sec:preliminaries}) of the common constraint are equal for all players. Now, a set of strategies $\hQ(\underline{r}) = (Q_1(\underline{r}),...,Q_K(\underline{r}))$ is a NoE if it satisfies the first order conditions for all $ k = 1,2,...,K$:
\begin{align}
& \frac{\partial L_k^{R}}{\partial Q_k}  = r_k \frac{\partial v_k(Q_k,Q_{-k})}{\partial Q_k} - \lambda I  + M_k  = 0 \nonumber \\
& M_k \succeq 0, ~ \mbox{Tr} [ M_k Q_k ] = 0 \nonumber \\
& \lambda > 0, ~  \sum_{i=1}^{K} \mbox{Tr} \left [ Q_i \right ] - P_{tot}  = 0.
\label{e:Rosen_KKT}
\end{align}

\subsection{Relation between Pareto-optimal and NoE solutions}
Now we derive the relation between $\underline{\gamma}$ and $\underline{r}$ such that the solutions from obtained from the KKT conditions for the Pareto-optimal problem and the equilibrium problem are identical, \textit{i.e.,} $ Q^{*}(\underline{\gamma}) = \hQ(\underline{r}) $. Let us first consider the simple case when $Q^{*}_k \neq 0_{n_t \times n_t}$ for all $k = 1,2,...,K$. Defining $ \eta = \frac{\mu}{\lambda} $ and manipulating the KKT conditions, we see that for each $k = 1,2,...,K$,
\begin{equation}
\eta r_k \mbox{Tr} \left [ \frac{\partial v_k(Q_k,Q_{-k})}{\partial Q_k} Q_k \right ] = \sum_{i=1}^{K} \gamma_i \mbox{Tr} \left [ \frac{\partial v_i(Q_i,Q_{-i})}{\partial Q_k} Q_k \right ].
\label{e:pareto_rosen}
\end{equation}
Let us define $b_k = r_k \left . \mbox{Tr} \left [ \frac{\partial v_k(Q_k,Q_{-k})}{\partial Q_k} Q_k \right ] \right |_{Q^{*}} $, $\underline{b} = (b_1,b_2,...,b_k)^T$ and the elements of matrix $A$ as $ A_{ki} = \mbox{Tr} \left . \left [ \frac{\partial v_i(Q_i,Q_{-i})}{\partial Q_k} Q_k \right ] \right |_{Q^{*}}$. The above condition can be compactly written as $ \eta \underline{b} = A \underline{\gamma} $. As the weights $r_i$ are relative to each other, we assign $r_K = 1$ to constrain our problem. Assuming that $\underline{\gamma}$ and $Q^{*}$ are known, we can solve for $r_1,r_2,...,r_{K-1},\eta$.

\begin{example}
For the ADBC, evaluating the elements of $A$, we get 
\begin{equation}
A_{ki} = \left \{
    \begin{array}{cc}
     \text{Tr} \left [ (Q_1 + ... + Q_k + N_k)^{-1} Q_k \right ] &  \text{if} ~~  k = i \\
     \text{Tr}  [ ( (Q_1 + ... + Q_i + N_i)^{-1} - & \\
         (Q_1 + ... + Q_{i-1} + N_i)^{-1} )  Q_k  ] &  \text{if} ~~ k < i \\
     0 & \text{if} ~~ k > i   \end{array} \right.
\label{e:A_matrix_ADBC}
\end{equation}

For the regime of $\underline{\gamma}$ in which $Q^{*}_k \neq 0_{n_t \times n_t} $ for all $k$, it is clear from (\ref{e:A_matrix_ADBC}) that $A$ is an upper-triangular matrix with positive diagonal elements and negative-off diagonal elements. Clearly, the structure of the diagonal and off-diagonal elements ensures that 
the inverse of the matrix is upper-triangular with all non-negative elements and hence for every weight vector $\underline{r}$, we can derive the set of weights $\underline{\gamma}$ which results in the same $Q^{*}$. This shows that every equilibrium point of the ADBC is Pareto-optimal.   

Now, let us consider the scenario in which $Q^{*}_k = 0_{n_t \times n_t}$ for some $k$. This implies that no transmission is scheduled for the $k^{th}$ user and this user does not enter the game for the considered value of $\underline{\gamma}$. In addition, we note that (\ref{e:pareto_rosen}) reduces to the trivial equation $ 0 = 0 $.  For such a scenario, we define a new BC with $K-1$ users by eliminating the $k^{th}$ user. Clearly, the newly defined BC is an ADBC with $Q^{*}_i  \neq 0, ~ \forall i $ and hence the above procedure can be used to find the weights to enforce the Pareto-optimal solution. If all the players except one have $Q^{*}_i = 0_{n_t \times n_t} $ we have only one player remaining and the problem reduces to a degenerate game. 
\end{example}
Similarly, exploiting the structure of the utility functions in (\ref{e:BC_util_2}) and (\ref{e:SPMAC_util}) (DPC encoding for the BC and SIC for the sum power MAC), it can be shown that for the general Gaussian BC and the sum power MAC, the equilibria corresponding to a given decoding order lie on the Pareto-optimal boundary of the rate region achievable by that particular encoding/decoding order. We formalize the above discussion into the following theorem.

\begin{thm}
Every NoE of a BC and sum power MAC, given the corresponding encoding/decoding order, lie on the Pareto-optimal boundary of the achievable rate region corresponding to that given encoding/decoding order.
\label{thm:Pareto-optimality}
\end{thm}

\subsection{Discussion}
\begin{itemize}
\item When the control variables of each player are scalars (and not matrices), the condition in (\ref{e:pareto_rosen}) reduces to
\begin{equation}
\eta r_k \frac{\partial v_k(Q_k,Q_{-k})}{\partial Q_k}  = \sum_{i=1}^{K} \gamma_i \frac{\partial v_i(Q_i,Q_{-i})}{\partial Q_k}.
\end{equation}
The above condition for the two-player game with scalar control variables has been derived in \cite{Jacek09} by computing the Lagrangian only w.r.t the common constraints and ignores the individual constraints on the control variables. While ignoring the individual constraints appears to give the same condition for the scalar scenario, we note that individual constraints can not be ignored in general. 
\item The similarity between the equilibrium characterization for the BC and the sum power MAC suggests that there exists a game-theoretic duality between the MAC and the BC, which we formalize in the rest of the paper. 
\end{itemize}

\subsection{Two user Example}
We consider a two-user single antenna ADBC to illustrate the solution and provide some insights into the relation between the weights characterizing equilibria and Pareto-optimal solutions. For a two-user one-antenna ADBC $(n_t = n_1 = ... = n_K = 1)$ with $Q_k$ representing the power allocated (instead of a covariance matrix) to the $k^{th}$ user. Assuming $N_1 < N_2$, the Pareto-optimal utility for the first and second users is given as $v_1(Q_1,Q_2) = \mbox{log}(Q_1 + N_1) - \mbox{log}(N_1) $ and $ v_2(Q_1,Q_2) = \mbox{log}(Q_1 + Q_2 + N_2) - \mbox{log}(Q_1 + N_2) $ respectively with $Q_1 + Q_2 = P_{tot}$ where $P_{tot}$ is the total power available at the transmitter. Substituting the utility functions in (\ref{e:pareto_rosen}), the relation between the two sets of weights can be derived as:
\begin{equation}
\frac{\gamma_1}{\gamma_2} = \frac{r_1}{r_2} + \frac{(Q_1 + N_1) (Q_2)}{ (Q_1 + Q_2 + N_2) (Q_1 + N_2)}.
\label{e:2user}
\end{equation}

\begin{figure}
\centering{
    \includegraphics[width=3in,height=2.5in]{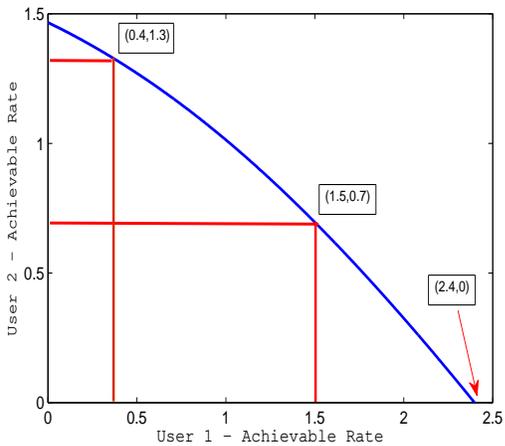}
    \caption{Plot showing the Pareto optimal frontier for a 2-user ADBC and several normalized equilibrium points}
    \label{fig:NoE}}
\end{figure}
Fig. \ref{fig:NoE} shows the Pareto-frontier for the 2-user ADBC with $P_{tot} = 10$, $ N_1 = 1$ and $N_2 = 3$ respectively. We first note that each point not on the axes of this curve can be characterized by a unique value of $\underline{\gamma}$. However several weight vectors result in the same Pareto-optimal boundary when this point is on the axes. For example, all $\gamma_1 \in [11/24,1]$ result in the same Pareto-optimal point $(v_1,v_2) = (2.4,0)$.

In Fig.\ref{fig:NoE}, the rate tuples for $\gamma_1 = 0.41$ and $\gamma_1 = 0.375$ are marked and given by $(v_1,v_2) = (1.5,0.7)$ and $(v_1,v_2) = (0.4,1.3)$ respectively. Using (\ref{e:2user}), we can show that these correspond to a $r_1/r_2 = 0.35$ and $r_1/r_2 = 0.11$  respectively. From Proposition 1, we infer that the the equilibrium point $r_1/r_2 = 0.35 > 0.25$ is the unique NoE of the game for $\gamma_1 = 0.41$. Thus, a regulator can use these weights to enforce the equilibrium rate $(1.5,0.7)$ using the taxation method described in Section \ref{sec:preliminaries}. In contrast, we see that for the rate tuple corresponding to $r_1/r_2 = 0.11 < 0.25 $, the corresponding NoE may not be a unique one. Finally, for $(v_1,v_2) = (2.4,0)$, we note that there is one active player and hence is a degenerate game. However, using (\ref{e:2user}), we see that by choosing any ratio $r_1/r_2 = \gamma_1/\gamma_2 > 11/13 $ the regulator can try to impose the fact that only the stronger user is allocated all the power for this game.

%% file: duality.tex
In this section, we establish a \emph{game-theoretic duality} between the MAC and BC by exploiting the properties of the signal transformation developed for the information theoretic duality in \cite{Vishwanath03}. We believe that such a transformation can provide new techniques for the computation of GNEs and NoEs for the GNEP. We begin by establishing the relationship between the equilibria of the MAC and the sum power MAC by exploring the relationship between an NEP and a suitably defined GNEP.

\subsection{Relationships between NEP and GNEPs}
Let $P$ be the total power available to all the users in a system. Define a NEP with $K$ players, utility functions $ v_k(Q_k,Q_{-k}) $ such that $v_k$ is concave in $Q_k$ and continuous in $Q_{-k}$, and the feasible strategy of each player as given below (Note that the presentation of this section, while tuned towards showing a duality between MAC and BC, can be extended to constraints and utility functions of a more general nature).
\begin{eqnarray}
(N1)         &&  \arg \max_{Q_k} ~~ v_k(Q_k,Q_{-k})  \nonumber \\
                 &&  Q_k \succeq 0, ~~ \mbox{Tr} [ Q_k ] \leq P_k.
\end{eqnarray}
In addition, we consider NEPs for which the power constraints satisfy the condition $ \sum_{k=1}^{K} P_k = P $. Note that this constraint is not inherent to the NEP. It just signifies the fact that we are interested in all such NEPs whose individual power constraints for each player sum up to $P$.

Let us now define a GNEP which is very closely related to $(N1)$.
\begin{eqnarray}
(G1)  &&  \arg \max_{Q_k} ~~ v_k(Q_k,Q_{-k})  \nonumber \\
                &&  Q_k \succeq 0, ~~ \sum_{k=1}^{K} \mbox{Tr} [ Q_k ] \leq P.
\end{eqnarray}
Observe that $(G1)$ is a GNEP with jointly convex constraints. Clearly, both the games $(N1)$ and $(G1)$ have the same number of players, with each player having the same utility function. The key difference between the two problems is that while the strategy set of each player is independent of the strategies of other players for $(N1)$, they are dependent on the strategies of other players from $(G1)$. In addition, we note that the feasible set of $(N1)$ is always a subset of the feasible set of $(G1)$. We now discuss the relationship between GNEs of $(G1)$ and the NEs of $(N1)$.

\begin{figure}
\centering{
\includegraphics[width=3.25in,height=2.5in]{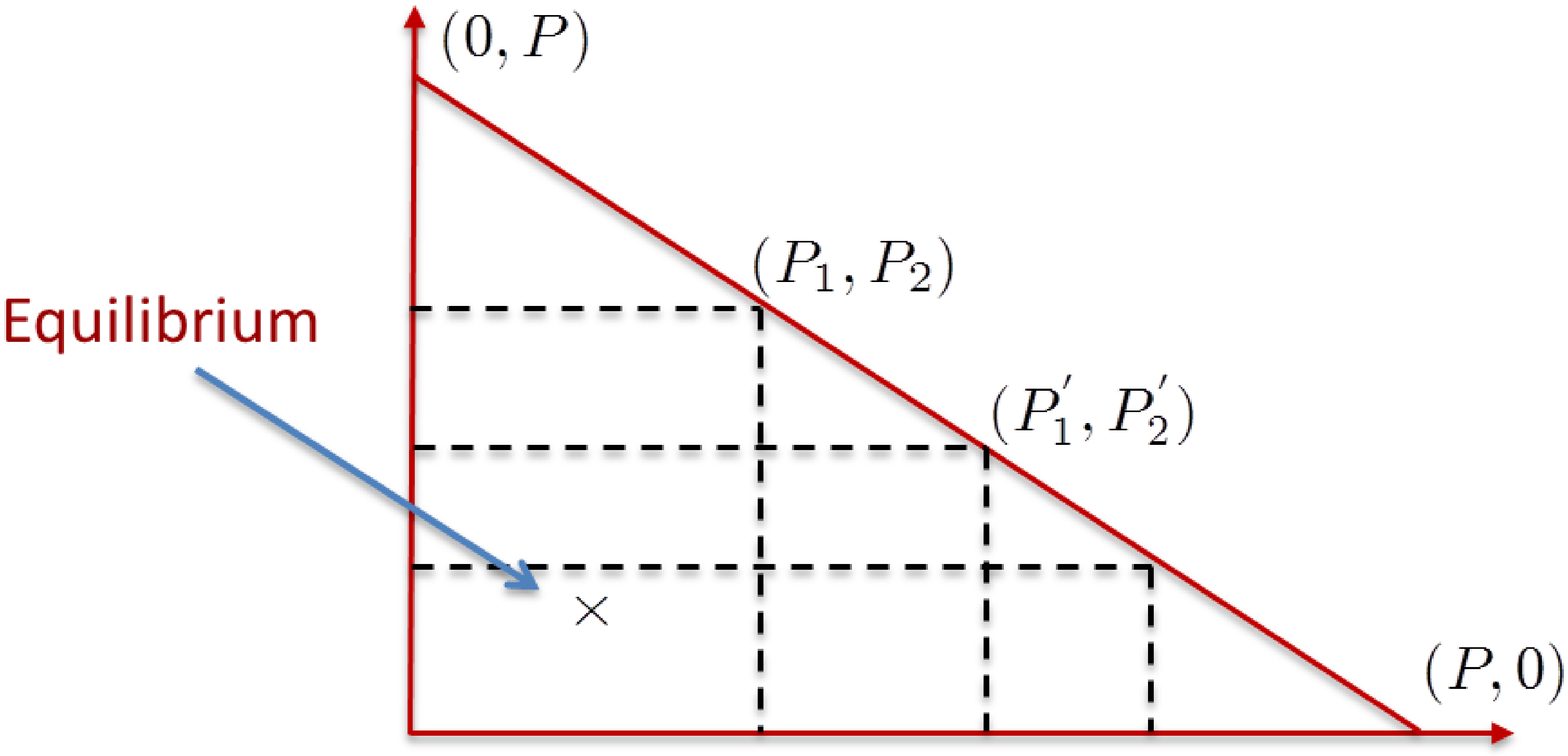}
\caption{Figure showing the relationship between the equilibria for $(N1)$ and $(G1)$ for scalar valued strategies in a $2$ player game.}
\label{fig:NEP-GNEP}}
\end{figure}

\begin{prop}
A vector of strategies $Q^{*}$ is a GNE of $(G1)$ if and only if $Q^{*}$ is a NE of all problems $(N1)$ for which $Q^{*}$ is in the feasible set of strategies.
\label{prop:NE_GNE}
\end{prop}
In other words, consider all the NEPs for which $Q^{*}$ is a member of the set of strategies, \emph{i.e.,} $ \mbox{Tr} [Q^{*}_k] \leq P_k $ with the exogenous constraint that $\sum_{k=1}^{K} {P_k} = P$. Then $Q^{*}$ is a NE for all such NEPs if and only if $Q^{*}$ is a GNE of $(G1)$.
\begin{IEEEproof}
Let us begin by assuming that $Q^{*}$ is a GNE of the jointly convex GNEP defined by $(G1)$. Then,
\begin{equation}
v_k(Q^{*}_k,Q^{*}_{-k}) \geq v_k(Q_k,Q^{*}_{-k}),
\end{equation}
for all $Q_{k} \in \Psi_k(Q_{-k}) $. Now let $(P^{'}_1,P^{'}_2,...,P^{'}_K)$ be a vector such that $ 0 \leq \mbox{Tr}[Q^{*}_k] \leq P^{'}_k $ with $ \sum_{k=1}^{K} P^{'}_k = P$ and consider the NEP:
\begin{eqnarray}
(N2)   &&  \arg \max_{Q_k} ~~ v_k(Q_k,Q_{-k})  \nonumber \\
                 &&  Q_k \succeq 0, ~~ \mbox{Tr} [ Q_k ] \leq P^{'}_k.
\end{eqnarray}
It is straightforward to observe that the feasible set of $(N2)$ is a subset of the feasible set of $(G1)$. Using the fact that $Q^{*}$ is a GNE of $(G1)$, it is easy to observe that
\begin{equation}
v_k(Q^{*}_k,Q^{*}_{-k}) \geq v_k(Q_k,Q^{*}_{-k}), \forall ~ 0 \leq \mbox{Tr}[Q_{k}] \leq P^{'}_k,
\end{equation}
and hence $ Q^{*} $ is a NE of $(N2)$. As the vector $(P^{'}_1,P^{'}_2,...,P^{'}_K)$ has been chosen arbitrarily, it is clear that if $Q^{*}$ is a GNE of $(G1)$ it is also a NE for all NEPs $(N1)$ for which it is a member of the feasible set.

Next, we assume that $Q^{*}$ is a NE of all the problems of the type $(N1)$ for which it is an element of the feasible set and suppose that $Q^{*}$ is \emph{not} a GNE of $(G1)$. Then, there exists a $m$ and $\tQ_m$ such that
\begin{equation}
v_m(Q^{*}_m,Q^{*}_{-m}) < v_m(\tQ_m,Q^{*}_{-m}).
\end{equation}
Let $ \mbox{Tr}[\tQ_m] = {\tilde P}$. Now, consider a vector of the form $(P^{'}_1,...P^{'}_{m-1},{\tilde P},P^{'}_{m+1},...,P^{'}_K)$ such that $\sum_{k=1,k \neq m}^{K} P^{'}_k + {\tilde P} = P$ and a NEP given by
\begin{eqnarray}
(N3)   && \arg \max_{Q_k} ~~ v_k(Q_k,Q_{-k})  \nonumber \\
           &&  Q_k \succeq 0, ~~ \mbox{Tr} [ Q_k ] \leq P^{'}_k, k \neq m \nonumber \\
           &&  Q_k \succeq 0, ~~ \mbox{Tr} [ Q_m ] \leq {\tilde P}
\end{eqnarray}
As the feasible set of $(N3)$ is a subset of the feasible set of $(G1)$, it is clear that
\begin{equation}
v_m(Q^{*}_m,Q^{*}_{-m}) < v_m(\tQ_m,Q^{*}_{-m}).
\end{equation}
for $(N3)$. This contradicts the fact that $Q^{*}$ is a NE of any problem of type $(N1)$ for which it is a member of the feasible set and proves the proposition.
\end{IEEEproof}

\begin{figure}
\centering{
\includegraphics[width=3.25in,height=2.5in]{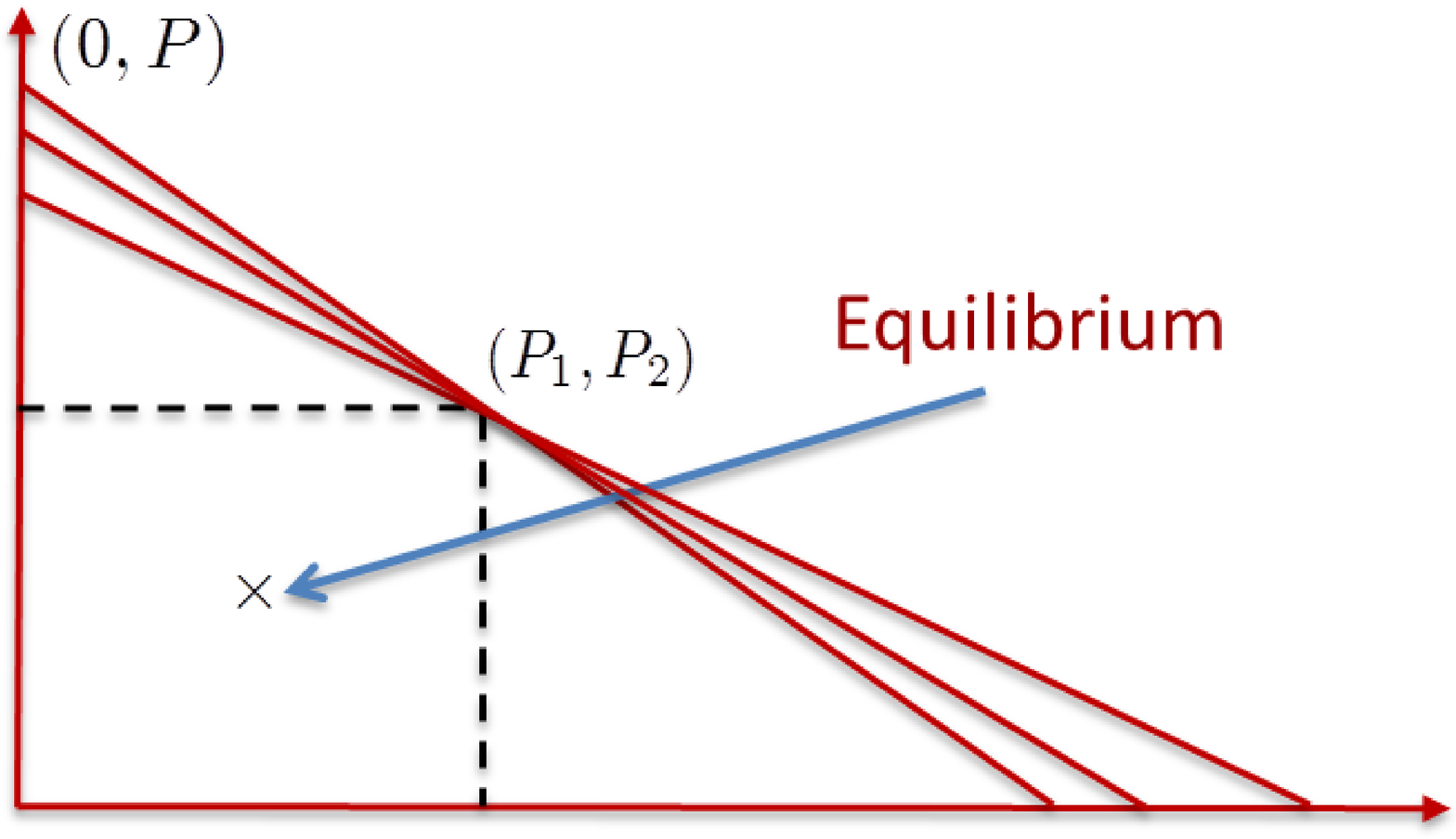}
\caption{Figure showing the relationship between the equilibria for $(N1)$ and $(G2)$ for scalar valued strategies in a 2 player game.}
\label{fig:GNEP-NEP}}
\end{figure}

\begin{prop}
If for every positive weight vector $\underline{\alpha}= (\alpha_1,\alpha_2,...,\alpha_K)$,  $Q^{*}$ is a GNE for a GNEP of the form,
\begin{eqnarray}
(G2) && \arg \max_{Q_k}   v_k(Q_k,Q_{-k}) \nonumber \\
  && Q_k \succeq 0, \sum_{k=1}^{K} \frac{ \textrm{Tr} [Q_k] }{\alpha_k} \leq \sum_{k=1}^{K} \frac{P_k}{\alpha_k},
\end{eqnarray}
then $Q^{*}$ is a NEP of $(N1)$.
\label{prop:GNE_NE}
\end{prop}
\begin{IEEEproof}
Define $(G3)$ which contains the common feasible set of all the GNEPs for every value of $\underline{\alpha}$:
\begin{eqnarray}
(G3)  && \arg \max_{Q_k}  v_k(Q_k,Q_{-k}), ~ Q_k \succeq 0 \nonumber \\
  && \bigcup \limits_{\underline{\alpha} > 0 } \left \{  \sum_{k=1}^{K} \frac{ \textrm{Tr} [Q_k] }{\alpha_k} \leq \sum_{k=1}^{K} \frac{P_k}{\alpha_k} \right \}.
\end{eqnarray}
We first note that the feasible set of $(G3)$ is an intersection of infinitely many convex sets and hence is a convex set itself. Clearly, $(G3)$ is jointly convex. It can be easily shown that the feasible set of $(G3)$ is identical to the feasible of $(N1)$. Thus, we observe that if $Q^{*}$ is a GNEP of the $(G3)$ for every $\underline{\alpha}$, then $Q^{*}$ is also a GNEP to $(N1)$ as both the games have the same utility functions for the players and the same feasible sets.  
\end{IEEEproof}

We now utilize the relationship between the equilibria of $(G1)$ and $(N1)$ to first derive the relation between the equilibria of the MAC and the sum power MAC. Substituting the utility functions $v_k(Q_k,Q_{-k})$ from (\ref{e:SPMAC_util}) in $(N1)$ and $(G1)$, we infer from Proposition \ref{prop:NE_GNE} that every GNE of the sum power MAC is also a NE of a MAC with appropriately defined power constraints. From Proposition \ref{prop:GNE_NE}, we then infer that every NE of a MAC is also a GNE of all scaled sum power MACs (where the feasible set scaling is as discussed in Proposition \ref{prop:GNE_NE}). Thus, we have established a relationship between the Nash equilibria of a MAC and the generalized Nash equilibria of a sum power MAC. Note that this result is true for any given decoding order and hence applies to all possible decoding orders. As a next step towards deriving the duality between the MAC and the BC, we now investigate the properties of GNEs when GNEPs are transformed using a special linear transformation.

\subsection{Relationship between two different GNEPs}
The information theoretic duality between the MAC and the BC \cite{Vishwanath03} is based on a transformation from the variables from the sum power MAC to the BC (and vice versa) such that the utilities are preserved with the transformation (see Equations (11) and (15) in \cite{Vishwanath03} for the transformation equation). We now explore the relationship between the equilibria of two GNEPs related by a similar linear transformation.
\begin{prop}
Let us consider two GNEPs (G4) and (G5) defined as follows:
\begin{eqnarray}
(G4) ~~ \arg \max_{Q_k} v_k(Q_k,Q_{-k}) \nonumber \\
     Q_k \succeq 0, \sum_{k=1}^{K} \mbox{Tr}[Q_k] \leq P,
\end{eqnarray}
\begin{eqnarray}
(G5)~~ \arg \max_{S_k} u_k(S_k,S_{-k}) \nonumber \\
     S_k \succeq 0, \sum_{k=1}^{K} \mbox{Tr}[S_k] \leq P,
\end{eqnarray}
and satisfying the following properties.
\begin{itemize}
\item $v_k$ is concave in $Q_{k}$ and continuous in $Q_i$ for all $i=1,...,K$.
\item $u_k$ is concave in $S_{k}$ and continuous in $S_i$ for all $i=1,...,K$.
\item There exist matrices $ A_k, B_k $ independent of $Q_k$ and $S_{k}$ (but may depend on $Q_{-k}$ and $S_{-k}$) such that for the linear transformations $S_k = A_k Q_k A^{H}_k $ and $Q_k = B_k S_k B^{H}_k $, we have that $ v_k(Q_{k},Q_{-k}) = u_k(S_k,S_{-k})$ and
\item $ \sum_{k=1}^{K} \mbox{Tr}( A_k Q_k A^{H}_k ) \leq P$ and $ \sum_{k=1}^{K} \mbox{Tr}( B_k S_k B^{H}_k ) \leq P $.
\end{itemize}
Then, $ Q^{*} $ is a GNE of $(G4)$ if and only if $S^{*}$, whose $kth$ component is given as $ S^{*}_{k} = A_k Q^{*}_k A^{H}_k $, is a GNE of (G5) and $S^{*}$ is a GNE of (G5) if and only if $Q^{*}$, whose $kth$ component is given as $ Q^{*}_k = B_k S^{*}_k B^{H}_k $, is a GNE of (G4). In addition, the utility at equilibrium for both the games is identical, \textit{i.e.,} $ v_k(Q^{*}_k,Q^{*}_{-k}) = u_k(S^{*}_k,S^{*}_{-k}) $.
\label{prop:duality}
\end{prop}
\begin{IEEEproof}
We begin by assuming that $Q^{*}$ is a GNE of (G4). As $Q^{*}$ is a GNE  of a jointly convex game with utility $v_k(Q_k,Q_{-k})$, we have that
\begin{equation}
\mbox{Tr} \left [ \frac{\partial v_k}{\partial Q_k} \left |_{Q^{*}} \right ( Q_k - Q^{*}_k ) \right ] \geq 0,
\end{equation}
for all $ k=1,2,...,K$ and all feasible $Q_k$. Now, using the fact that $ Q_k = B_k S_k B^{H}_k $, we have that
\begin{equation}
q_{k,ij} = \sum_{m,n} b_{k,im} s_{k,mn} b^{*}_{k,jn}. 
\end{equation}
Differentiating with respect to $s_{mn} $, we have that $ \frac{\partial  q_{k,ij} }{\partial s_{k,mn} } = b_{k,im} b^{*}_{k,jn} $. Now using the fact that $ v_k(Q_k,Q_{-k}) = u_k(S_k,S_{-k}) $ and that $ Q_k $ can be expressed as a linear function of $S_k$, we have that
\begin{align}
\frac{\partial u_k }{\partial s_{k,mn} } & = \frac{\partial v_k }{\partial s_{k,mn} } =  \sum_{i,j} \frac{\partial v_k }{\partial q_{k,ij} } \frac{\partial q_{k,ij} }{\partial s_{k,mn} } \nonumber \\
         & = \sum_{i,j} \frac{\partial v_k }{\partial q_{k,ij} } b_{k,im} b^{*}_{k,jn} \nonumber \\ 
				 & = \left [ b^{*}_{1n} ~ b^{*}_{2n} ~ \hdots  \right ]  
				\underbrace{\left [ \begin{array}{ccc} \frac{\partial v_k}{\partial q_{k,11}} & \frac{\partial v_k}{\partial q_{k,21}} & ... \\ 
				                           \frac{\partial v_k}{\partial q_{k,12}} & \frac{\partial v_k}{\partial q_{k,22}} & ... \\
																	 \vdots  & \ddots & \vdots \\ \end{array} \right ]}_{\frac{\partial v_k}{\partial Q_k}}
																	 \left [  \begin{array}{c} b_{k,1m} \\ b_{k,2m} \\ \vdots  \end{array} \right ] 
\end{align} 
Consolidating the above series of equations into a matrix, we get
\begin{equation}
\frac{\partial u_k }{\partial S_k } = B^{H}_k \frac{\partial v_k}{ \partial Q_k } B_k.
\end{equation}
Now using the fact that $Q^{*}$ is a GNE of $(G4)$, for all $k = 1,2, \hdots, K $, we have that 
\begin{eqnarray}
& ~ & \mbox{Tr} \left [ \frac{\partial u_k}{\partial S_k} \left |_{S^{*}} \right ( S_k - S^{*}_k ) \right ] \geq 0,	 \nonumber \\
& \Leftrightarrow & \mbox{Tr} \left [ B^{H}_k \frac{\partial v_k}{\partial Q_k} \left |_{Q^{*}} B_k \right ( Q_k - Q^{*}_k ) \right ] \geq 0, \nonumber \\
& \Leftrightarrow & \mbox{Tr} \left [ \frac{	\partial v_k}{\partial Q_k} \left |_{Q^{*}} \right ( Q_k - Q^{*}_k ) \right ] \geq 0.
\end{eqnarray}
This implies that if $Q^{*}$ is a GNE of $(G4)$, then $S^{*}$ is a GNE of $(G5)$. The converse can be shown in a similar manner, thus completing the proof. 
\end{IEEEproof}
Proposition \ref{prop:duality} shows that for any two GNEPs which satisfy the given properties, the GNEs and the achieved utilities are identical. In other words, this provides a technique to transform one GNEP to another that might permit simpler analysis.

\subsection{Game based MAC-BC duality}
We now present a game-theory based dual relationship between the MAC and the BC. To this end, we first establish a relationship between the equilibria of the MAC and the sum power MAC and then connect the equilibria of the sum power MAC to the BC. 

\emph{Relationship between MAC and sum power MAC:} Let us first substitute the utility functions for the MAC from (\ref{e:SPMAC_util}) in Prop. \ref{prop:NE_GNE} which relates the GNEs of a GNEP to the NEs of a appropriately defined NEP. More precisely, Prop. \ref{prop:NE_GNE} states that a vector of strategies $Q^{*}$ is a GNE of the sum power MAC if and only if $Q^{*}$ is an NE for all MACs for which $Q^{*}$ is part of the feasible set of strategies. 

Now consider a MAC in which the $k^{th}$ player has a power constraint $\mbox{Tr}[Q_k] \leq P_k $. We know from \cite{Srinivas13MAC} that the NE is achieved at $ \mbox{Tr}[Q_k] = P_k$. In other words, given a decoding order for the transmitters, the NE strategy of the MAC is to transmit at the highest power. It is easy to see that $Q^{*}$ is a point on the line $ \sum_{k=1}^{K} Tr[Q_k] = P $ and thus is not a member of the feasible set of strategies for any other problem of type $(N1)$ with $(P^{'}_1,...,P^{'}_K) \neq (P_1,P_2,...,P_K)$ (see Fig. \ref{fig:NEP-GNEP}). $Q^{*}$ is thus an NE for all problems of type $(N1)$ for which it is a member of the feasible set of strategies. Thus from Proposition \ref{prop:NE_GNE}, we infer that $Q^{*}$ is also a GNE of the sum power MAC. This gives our first result which states that the GNEs of a sum power MAC can be constructed from the NEs of all the MACs which satisfy the exogenous power constraint of the sum power MAC. 

Next we consider a GNE $Q^{*}$ of the sum power MAC. From Theorem	 \ref{thm:Pareto-optimality} in Section \ref{sec:relationship}, we know that $Q^{*}$ achieves a point of the Pareto-boundary of the rate region for the sum power MAC. Using Prop. \ref{prop:NE_GNE} it is easy to see that $Q^{*}$ also an NE for the MAC with individual power constraints given by $ P^{*}_k = \mbox{Tr}[Q_k] $. To summarize, \emph{we have now shown that the set of GNEs of a sum power MAC is the union of the set of NEs of the corresponding MACs.} 

Finally, Prop. \ref{prop:GNE_NE} shows that if $Q^{*}$ is a GNE for all sum power MACs which satisfy power constraints of the form $ \sum_{k=1}^{K} \frac{Tr[Q_k]}{\alpha_k} \leq \frac{P_k}{\alpha_k}$ for every positive weight vector $\underline{\alpha}$, then $Q^{*}$ is a NE of a MAC. In other words, \emph{the set of NEs of a MAC are the intersection of the set of all GNEs of corresponding sum power MACs.} 

\emph{Relationship between sum power MAC and BC:} Let us consider Prop. \ref{prop:duality} which provides for a way to transform one GNEP to another and relates the GNEs of such transformed GNEPs. We use this transformation to establish the relationship between the GNEs of the sum power MAC and the BC. Let $v_k$ and $u_k$ be the utility functions of the sum power MAC and the BC respectively. From equations (11) and (15) of \cite{Vishwanath03}, we know that there exist linear transformations $S_k = A_k Q_k A^{H}_k $ and $Q_k = B_k S_k B^{H}_k $, with $A_k$ and $B_k$ independent of $Q_{k}$ and $S_{k}$ respectively, satisfying the power constraints, and which transform the utility function of the sum power MAC (same as the utility function for the MAC) into the utility function of the BC. Thus, from Proposition \ref{prop:duality}, it is clear that for every GNE of the sum power MAC there exists an GNE of the BC and vice versa.  

Now combining the results from Propositions \ref{prop:NE_GNE} and \ref{prop:duality}, we can infer that for every equilibrium point of the MAC there exists an equilibrium point of the BC and vice versa. We summarize this result in the following theorem.

\begin{thm}
For a given decoding order for a MAC (and corresponding encoding for the BC), the set of all NEs of the MAC is the set of all GNEs common to the corresponding (as defined in $(G2)$ and $(G3)$ ) BCs and the set of all GNEs of the BC is the union of the set of NEs of the corresponding (as defined in $(N1)$ and $(N2)$) MAC. In other words, the equilibrium rate regions of the MAC and the BC are identical. 
\end{thm}

This duality allows us to easily find the GNEs of the BC providing a method to transform the equilibria of the MAC to the BC. In general, a game-theoretic duality between the MAC and the BC could be potentially extended to the scenario with multiple constraints at the transmitter and allow for the translation of several equilibrium results for the MAC to the BC.

%% file: conclusions.tex
In this paper, we proposed a game theoretic model for the Gaussian BC and a related problem, the MAC channel with sum power constraints. By modeling both scenarios as a generalized Nash equilibrium problems, we characterized the existence and uniqueness of normalized equilibrium achieving strategies and utilities. We then proposed a characterization for Pareto-efficient equilibria and show that every point on the Pareto-optimal boundary of the BC and sum power MAC for a given decoding order is a normalized equilibrium corresponding to a defined parametrization. Using the proposed characterization, a regulator can implement a desired Pareto-optimal solution as the solution to the game between receivers for the broadcast channel and transmitters for the sum power MAC. Next, we establish a several relationships between the equilibria of the MAC, sum power MAC and the BC and derive a game-theoretic duality between the MAC and the BC which shows that for a given decoding/encoding order, the equilibria of the sum power MAC and the BC are identical. Thus, the equilibrium rate region of the MAC is identical to the equilibrium rate region of the BC, thus demonstrating a information theory based technique to evaluate the GNEs of the BC.

%% file: Broadcast_Games.bbl
\begin{thebibliography}{10}
\providecommand{\url}[1]{#1}
\csname url@samestyle\endcsname
\providecommand{\newblock}{\relax}
\providecommand{\bibinfo}[2]{#2}
\providecommand{\BIBentrySTDinterwordspacing}{\spaceskip=0pt\relax}
\providecommand{\BIBentryALTinterwordstretchfactor}{4}
\providecommand{\BIBentryALTinterwordspacing}{\spaceskip=\fontdimen2\font plus
\BIBentryALTinterwordstretchfactor\fontdimen3\font minus
  \fontdimen4\font\relax}
\providecommand{\BIBforeignlanguage}[2]{{%
\expandafter\ifx\csname l@#1\endcsname\relax
\typeout{** WARNING: IEEEtran.bst: No hyphenation pattern has been}%
\typeout{** loaded for the language `#1'. Using the pattern for}%
\typeout{** the default language instead.}%
\else
\language=\csname l@#1\endcsname
\fi
#2}}
\providecommand{\BIBdecl}{\relax}
\BIBdecl

\bibitem{Scutari10}
G.~Scutari, D.~P. Palomar, F.~Facchinei, and J.~S. Pang, ``Convex optimization,
  game theory, and variational inequality theory,'' \emph{IEEE Signal
  Processing Magazine}, pp. 35--49, May 2010.

\bibitem{survey11}
K.~Akkarajitsakul, E.~Hossain, D.~Niyato, and D.~I. Kim, ``Game theoretic
  approaches for multiple access in wireless networks: A survey,'' \emph{IEEE
  Communications Surveys and Tutorials}, vol.~13, no.~3, pp. 372--395, 2011.

\bibitem{La04}
R.~J. La and V.~Ananthram, ``A game-theoretic look at the {G}aussian multiple
  access channel,'' \emph{DIMACS Series in Discrete Mathematics and Theoretical
  Computer Science}, vol.~66, no.~4, pp. 87--106, 2004.

\bibitem{Lai08}
L.~Lai and H.~El~Gamal, ``The water-filling game in fading multiple-access
  channels,'' \emph{IEEE Transactions on Information Theory}, vol.~54, no.~5,
  pp. 2110 --2122, may 2008.

\bibitem{Belmega10}
E.~V. Belmega, S.~Lasaulce, M.~Debbah, M.~Jungers, and J.~Dumont, ``Power
  allocation games in wireless networks of multi-antenna terminals,''
  \emph{Telecommunication Systems}, vol.~47, pp. 109--122, 2011.

\bibitem{Srinivas11ISIT}
S.~Yerramalli, R.~Jain, and U.~Mitra, ``Coalition games for transmitter
  coooperation in wireless networks,'' in \emph{International Symposium on
  Information Theory}, Aug 2011, pp. 2706--2710.

\bibitem{Pang08intfchannel}
J.-S. Pang, G.~Scutari, F.~Facchinei, and C.~Wang, ``Distributed power
  allocation with rate constraints in {G}aussian parallel interference
  channels,'' \emph{IEEE Transactions on Information Theory}, vol.~54, no.~8,
  pp. 3471--3489, Aug 2008.

\bibitem{Rosen65}
J.~B. Rosen, ``Existence and uniqueness of equilibrium points for concave
  {N}-person games,'' \emph{Econometrica: Journal of the Econometric Society},
  vol.~33, no.~3, pp. 520--534, July 1965.

\bibitem{Su08}
Y.~Su and M.~Van Der~Schaar, ``A simple characterization of strategic behaviors
  in broadcast channels,'' \emph{Signal Processing Letters, IEEE}, vol.~15, pp.
  37--40, 2008.

\bibitem{Srinivas13MAC}
S.~Yerramalli, R.~Jain, and U.~Mitra, ``Coalitional games for transmitter
  cooperation in {MIMO} multiple access channels,'' 2013, url:
  http://arxiv.org/abs/1206.3350 (to appear in IEEE Transactions of Signal
  Processing, pending minor revision).

\bibitem{Pantisano2012spectrum}
F.~Pantisano, M.~Bennis, W.~Saad, and M.~Debbah, ``Spectrum leasing as an
  incentive towards uplink macrocell and femtocell cooperation,''
  \emph{Selected Areas in Communications, IEEE Journal on}, vol.~30, no.~3, pp.
  617--630, 2012.

\bibitem{Xu12}
C.~Xu, L.~Song, Z.~Han, Q.~Zhao, X.~Wang, and B.~Jiao, ``Interference-aware
  resource allocation for device-to-device communications as an underlay using
  sequential second price auction,'' in \emph{Communications (ICC), 2012 IEEE
  International Conference on}, 2012, pp. 445--449.

\bibitem{Akkarajitsakul2012mode}
K.~Akkarajitsakul, P.~Phunchongharn, E.~Hossain, and V.~K. Bhargava, ``Mode
  selection for energy-efficient d2d communications in lte-advanced networks: A
  coalitional game approach,'' in \emph{Communication Systems (ICCS), 2012 IEEE
  International Conference on}.\hskip 1em plus 0.5em minus 0.4em\relax IEEE,
  2012, pp. 488--492.

\bibitem{Zhang2009weighted}
L.~Zhang, Y.~Xin, and Y.-C. Liang, ``Weighted sum rate optimization for
  cognitive radio mimo broadcast channels,'' \emph{Wireless Communications,
  IEEE Transactions on}, vol.~8, no.~6, pp. 2950--2959, 2009.

\bibitem{Weingarten06}
H.~Weingarten, Y.~Steinberg, and S.~Shamai, ``The capacity region of the
  {G}aussian multiple-input multiple-output broadcast channel,'' \emph{IEEE
  Transactions on Information Theory}, vol.~52, no.~9, pp. 3936 --3964, Sept
  2006.

\bibitem{Vishwanath03}
S.~Vishwanath, N.~Jindal, and A.~Goldsmith, ``Duality, achievable rates, and
  sum rate capacity of {G}aussian {MIMO} broadcast channels,'' \emph{IEEE
  Transactions on Information Theory}, vol.~49, no.~10, pp. 2658--2668, Oct
  2003.

\bibitem{Zhang2012gaussian}
L.~Zhang, R.~Zhang, Y.-C. Liang, Y.~Xin, and H.~V. Poor, ``On gaussian mimo
  bc-mac duality with multiple transmit covariance constraints,''
  \emph{Information Theory, IEEE Transactions on}, vol.~58, no.~4, pp.
  2064--2078, 2012.

\bibitem{Facchinei07}
F.~Facchinei and C.~Karzow, ``Generalized {N}ash equilibrium problems,''
  \emph{4OR: A Quarterly Journal of Operations Research}, vol.~5, pp. 173--210,
  2007.

\bibitem{Pang10CR}
J.-S. Pang, G.~Scutari, D.~P. Palomar, and F.~Facchinei, ``Design of cognitive
  radio systems under temperature-interference constraints: A variational
  inequality approach,'' \emph{Signal Processing, IEEE Transactions on},
  vol.~58, no.~6, pp. 3251--3271, 2010.

\bibitem{Jacek09}
J.~B. Krawczyk and M.~Tidball, ``How to use rosen's normalized equilibrium to
  enforce a socially desireable {P}areto efficient solution,'' in
  \emph{Proceedings of $15^{th}$ International Conference on Computing in
  Economics and Finance}, July 2009.

\bibitem{Boyd}
S.~Boyd and L.~Vandenberghe, \emph{Convex Optimization}.\hskip 1em plus 0.5em
  minus 0.4em\relax Cambridge University Press, 2009.

\bibitem{Belmega11ineq}
E.~V. Belmega, M.~Jungers, and S.~Lasaulce, ``A generalization of a trace
  inequality for positive semi-definite matrices,'' \emph{The Australian
  Journal of Mathematical Analysis and Applications}, vol.~7, no.~2, 2011.

\bibitem{Furuichi11ineq}
S.~Furuichi and M.~Lin, ``Refinements of the trace inequalities of {B}elmega,
  {L}asaulce and {D}ebbah,'' \emph{The Australian Journal of Mathematical
  Analysis and Applications}, vol.~7, no.~2, pp. 1--4, 2011.

\end{thebibliography}
